\documentclass[10pt,english]{article}
\usepackage{ae,aecompl}

\usepackage[T1]{fontenc}
\usepackage[latin9]{inputenc}
\usepackage{geometry}
\usepackage{wrapfig}
\usepackage{url}
\usepackage{amsmath}
\usepackage{graphicx}
\usepackage{amssymb}

\providecommand{\tabularnewline}{\\}

\newcommand{\lyxaddress}[1]{
\par {\raggedright #1
\vspace{1.4em}
\noindent\par}
}

\pdfoutput=1
\renewcommand{\vec}[1]{\mbox{\boldmath$#1$}}
\date{}

\usepackage[pdfauthor={Jeffrey M. Dick},pdftitle={Calculation of the relative metastabilities of proteins in subcellular compartments of Saccharomyces cerevisiae},pdfsubject={}]{hyperref}

\usepackage{babel}

\begin{document}

\title{Calculation of the relative metastabilities of proteins in subcellular
compartments of \emph{Saccharomyces cerevisiae}}

\author{Jeffrey M. Dick}

\maketitle

\lyxaddress{\begin{center}
Department of Earth and Planetary Science\\
University of California\\
Berkeley, CA 94720
\par\end{center}}

\emph{Background}: The distribution of chemical species in an open
system at metastable equilibrium can be expressed as a function of
environmental variables which can include temperature, oxidation-reduction
potential and others. Calculations of metastable equilibrium for various
model systems were used to characterize chemical transformations among
proteins and groups of proteins found in different compartments of
yeast cells. 

\emph{Results}: With increasing oxygen fugacity, the relative metastability
fields of model proteins (including isoforms of glutaredoxin and thioredoxin,
and compartmental proteomes) for major subcellular compartments go
as mitochondrion, endoplasmic reticulum, cytoplasm, nucleus. Compared
with experimental determination of redox potential (Eh) in these compartments,
the order of the endoplasmic reticulum and nucleus is swapped. In
a metastable equilibrium setting at relatively high oxygen fugacity,
proteins making up actin are predominant, but those constituting the
microtubule occur with a low chemical activity. Nevertheless, interactions
of the microtubule with other subcellular compartments are essential
in cell development. A reaction sequence involving the microtubule
and spindle pole proteins was predicted by combining the known intercompartmental
interactions with a hypothetical program of oxygen fugacity changes
in the local environment. In further calculations, the most-abundant
proteins within compartments generally occur in relative abundances
that only weakly correspond to a metastable equilibrium distribution.
However, physiological populations of proteins that form complexes
often show an overall positive or negative correlation with the relative
abundances of proteins in metastable assemblages. 

\emph{Conclusions}: This study explored the outlines of a thermodynamic
description of chemical transformations among interacting proteins
in yeast cells. Full correspondence of the model with biochemical
and proteomic observations was not apparent, but the results suggest
that these methods can be used to measure the degree of departure
of a natural biochemical process or population from a local minimum
in Gibbs energy.

\section*{Author Summary}

Part of a cell's expenditure of metabolic fuel is directed toward
the formation of proteins, including their synthesis and transport
to other compartments. Even when it is normalized to the lengths of
the proteins, the energy required for protein formation is not a constant,
but depends on the composition and environment of the protein. If
these energy differences are quantified, the relative abundances of
model proteins in metastable equilibrium can be calculated. The compositions
of these metastable assemblages depend on local environmental variables
such as oxygen fugacity, which is a scale for oxidation-reduction
potential in a system. I calculate the oxygen fugacities for equal
chemical activities of model proteins in intercompartmental interactions
and use the results to obtain model values of oxygen fugacity for
subcellular compartments. I show that a environmental gradient of
oxygen fugacity can potentially drive the formation of proteins in
a sequential order determined by their chemical compositions and Gibbs
energies. I also show that the relative abundances of proteins within
compartments and of those that form complexes have a dynamic range
that can be approximated in some metastable equilibrium assemblages.
These results provide theoretical constraints on the natural emergence
of spatial and temporal patterns in the distributions of proteins
and imply that work done by maintaining oxidation-reduction gradients
can selectively alter the degree of formation of proteins and complexes.

\section*{Introduction}

Subcellular compartmentation is a basic feature of eukaryotic life
\cite{AlH95,Aw00,MR03,HGJ06}. There exist in eukaryotic cells gradients
between subcellular compartments of chemical properties such as $\mathrm{pH}$
\cite{PMJ89,IO95,LMM+98,WLA+00}, oxidation-reduction or redox state
\cite{HSL92,HAO+04,TG05,SCC+07} and chemical activity of water \cite{MKO+96,Gar00,VB05,PZK+95}.
Furthermore, the proteins required by yeast and other organisms are
unevenly localized throughout cells \cite{NN94,CAP+97,AOR98,CE99,KAH+02,HFG+03,PSL+07}.
Even within compartments or among the proteins that interact to form
complexes, the relative abundances or levels of different proteins
are not equal \cite{RAS+00,GPS00,PSC+04}, and different proteins
predominate in the various subcellular populations depending on growth
state of the cell \cite{KSH+08,HLG94}, and exposure to environmental
stress \cite{GSK+00,CRK+01,CTM+03,GLL+07}.

Much attention has been given to the use of thermodynamics in describing
and understanding driving forces in biological evolution. Energy minimization
imparts a direction for spontaneous change of a system, and response
of a system in this direction can at times be tied to an increase
in relative fitness \cite{SH05c,AH96,JF04,DZ07,Rei07}. A biological
system that moves away from minimum energy does not break the laws
of thermodynamics but couples its endergonic reactions with the exchange
of matter and energy in its surroundings \cite{Bat92,FH06,Wac88,SS98b}.
The thermodynamic characteristics of open systems are thus of particular
interest to biological evolution \cite{ME74,SK94,Wic79}; in particular,
the interactions of organisms with their environments are important
influences on the stable compositions and distributions of genes or
organisms \cite{Wri04,FJP05b,Wic80,LLN08}.

Why are proteins not equally distributed inside cells? Physical separation
of key enzymes is thought to be essential in the cytoskeletal network
and in regulation of metabolic pathways and other cellular functions
\cite{Sch85,GWJ+98,DMT05}. The patterns of subcellular structure
persist even though populations of proteins turnover through continual
degradation and synthesis in cells \cite{SRR39,Swi58,Hal58,BTS07},
and despite the endergonic, or energy-consuming, qualities of protein
biogenesis \cite{BD40,Mor78}. It can be shown that the relative abundances
of amino acids in proteins correlate inversely with the metabolic
cost of synthesis of the amino acid \cite{Sel03,Swi07}, which is
a temperature-dependent function \cite{ZBS07}. The starting premise
of this study, then, is that protein formation reactions are unfavorable
to different degrees, depending on the environments and compositions
of the biomolecules. 

The application of equilibrium chemical thermodynamics as a way to
characterize the relative stabilities of minerals as a function of
temperature, pressure and oxidation-reduction potential \cite{Kor66,ES67,Hel70c},
or to calculate the relative abundances of coexisting inorganic \cite{HSS95,See01}
and/or organic species \cite{SS98b,HRM+08}, is well documented in
the geochemical literature. An advantage of performing quantitative
chemical thermodynamic calculations for many different model systems
is that the equilibrium state serves as a frame of reference for describing
both reversible and irreversible chemical changes. For example, the
weathering of igneous rocks is an overall irreversible process but
the sequences of minerals formed can nevertheless be predicted after
initial formulation of the relative stability limits of the chemical
species involved \cite{Hel74,Hel79}. One of the motivations for this
study is to see whether a similar approach could be used to describe
the sequence of events in irreversible subcellular processes. 

The thermodynamic calculations reported in this study are based on
algorithms for calculating the standard molal Gibbs energies of ionized
proteins \cite{DLH06} and a chemical reaction framework that is used
to compute metastable equilibrium relative abundances of proteins
\cite{Dic08a}. The Supporting Information for this paper includes
the software package (Text S1) and the program script and data files
(Text S2) used to carry out these calculations. The theoretical approach
adopted here is based on the description of a chemical system in terms
of intensive variables. These variables are temperature, pressure
and the chemical potentials of the system. It is convenient to denote
the chemical potentials by the chemical activities or fugacities of
basis species, for example the activity of $\mathrm{H^{+}}$ (which
defines $\mathrm{pH}$) or the fugacity of oxygen. This permits comparison
of the parameters of the model with reference systems described in
experimental and other theoretical biochemical studies.

A few notes on terminology follow. \emph{Formation} of a protein refers
to the overall process of protein biosynthesis and translocation to
a specific compartment. \emph{Activity} and \emph{species} denote,
respectively, chemical activity and chemical species, not enzyme activity
or biological species. In the present study, activity coefficients
are taken to be unity, so the chemical activities are equivalent to
molal concentrations. Below, \emph{oxidation-reduction potential}
and oxygen fugacity are used synonymously, and \emph{redox} refers
specifically to $\mathrm{Eh}$. The oxidation-reduction potential
of a system can be expressed in terms of $\mathrm{Eh}$ using an equation
given in the Methods. The overall compositions of proteins in compartments
are referred to here as \emph{proteologs} (or model proteologs). The
\emph{interactions} of proteins are processes in which the proteins
come into physical contact, for example in transport processes between
compartments and in the formation of complexes. If a process results
in a change in the composition of a population of interacting proteins,
then a chemical \emph{reaction} has occurred. Protein-protein interactions
do not necessarily correspond to chemical reactions. However, a population
of interacting proteins does chemically react if the turnover rates
of the proteins are not all the same or if, through evolution, the
genes coding for the proteins undergo different non-synonymous mutations.
Model systems consisting of interacting proteins are useful targets
for assessing the potential for chemical reactivity, which might occur
on evolutionary time scales longer than the physical interactions.

The purpose of this study is to quantify using a metastable equilibrium
reference state the responses of populations of model proteins for
different subcellular compartments of \emph{S. cerevisiae} to gradients
of oxidation-reduction potential\emph{.} There are two major parts
to this paper. In the first part, the reactions corresponding to intercompartmental
interactions between isoforms (or homologs) of particular enzymes
and between proteologs are quantified by calculating the oxygen fugacities
for equal chemical activities of the reacting proteins or proteologs
in metastable equilibrium. A ranking of relative metastabilities of
the proteologs is discussed. Specific known interactions between compartments
are considered in order to derive values of the oxygen fugacity within
compartments that best metastabilize the corresponding proteologs
relative to those of other compartments. Equal-activity values of
the oxygen fugacity in the reactions are used to predict a sequence
of formation of model proteologs in response to a temporal oxidation-reduction
gradient.

In the second part of this paper, the relative abundances of model
proteins in metastable equilibrium are calculated and compared with
measured abundances. The range of protein abundances in a metastable
equilibrium population often approaches that seen in experiments over
a narrow window of oxygen fugacity. Positive and negative correlations
between the calculated and experimental relative abundances are found
in some cases. Local energy minimization and its opposition in the
cellular demands for selectivity in protein formation are discussed
as possible processes leading to the observed patterns.

\section*{Results and Discussion}

\begin{table}
\begin{centering}
\caption{\label{table:homologs}Subcellular isoforms of glutaredoxin, thioredoxin
and thioredoxin reductase in yeast$^{\mathbf{a}}$.}
\begin{tabular}{lllllrrr}
\hline 
{\footnotesize Protein} & {\footnotesize SWISS-PROT} & {\footnotesize Location} & \multicolumn{1}{c}{{\footnotesize Length}} & {\footnotesize Formula} & {\footnotesize $\Delta G^{\circ}$} & {\footnotesize $Z$} & {\footnotesize $\overline{Z}_{\mathrm{C}}$}\tabularnewline
\hline
 &  &  &  &  &  &  & \tabularnewline
\multicolumn{5}{c}{{\footnotesize Glutaredoxin}} &  &  & \tabularnewline
{\footnotesize GLRX1} & {\footnotesize P25373} & {\footnotesize Cytoplasm} & {\footnotesize 110} & {\footnotesize $\mathrm{C_{549}H_{886}N_{146}O_{170}S_{4}}$} & {\footnotesize -4565} & {\footnotesize -5.8} & {\footnotesize -0.182}\tabularnewline
{\footnotesize GLRX2} & {\footnotesize P17695} & {\footnotesize Mitochondrion} & {\footnotesize 143} & {\footnotesize $\mathrm{C_{715}H_{1161}N_{181}O_{213}S_{5}}$} & {\footnotesize -5617} & {\footnotesize 0.1} & {\footnotesize -0.255}\tabularnewline
{\footnotesize GLRX3} & {\footnotesize Q03835} & {\footnotesize Nucleus} & {\footnotesize 285} & {\footnotesize $\mathrm{C_{1444}H_{2195}N_{371}O_{463}S_{10}}$} & {\footnotesize -12031} & {\footnotesize -24.5} & {\footnotesize -0.094}\tabularnewline
{\footnotesize GLRX4} & {\footnotesize P32642} & {\footnotesize Nucleus} & {\footnotesize 244} & {\footnotesize $\mathrm{C_{1226}H_{1910}N_{316}O_{389}S_{6}}$} & {\footnotesize -10276} & {\footnotesize -17.8} & {\footnotesize -0.140}\tabularnewline
{\footnotesize GLRX5} & {\footnotesize Q02784} & {\footnotesize Mitochondrion} & {\footnotesize 150} & {\footnotesize $\mathrm{C_{762}H_{1200}N_{196}O_{227}S_{6}}$} & {\footnotesize -5841} & {\footnotesize -6.1} & {\footnotesize -0.192}\tabularnewline
 &  &  &  &  &  &  & \tabularnewline
\multicolumn{5}{c}{{\footnotesize Thioredoxin}} &  &  & \tabularnewline
{\footnotesize TRX1} & {\footnotesize P22217} & {\footnotesize Cytoplasm} & {\footnotesize 102} & {\footnotesize $\mathrm{C_{502}H_{785}N_{123}O_{150}S_{5}}$} & {\footnotesize -3969} & {\footnotesize -3.1} & {\footnotesize -0.211}\tabularnewline
{\footnotesize TRX2} & {\footnotesize P22803} & {\footnotesize Cytoplasm} & {\footnotesize 103} & {\footnotesize $\mathrm{C_{497}H_{780}N_{122}O_{153}S_{5}}$} & {\footnotesize -4056} & {\footnotesize -3.1} & {\footnotesize -0.197}\tabularnewline
{\footnotesize TRXB1} & {\footnotesize P29509} & {\footnotesize Cytoplasm} & {\footnotesize 318} & {\footnotesize $\mathrm{C_{1509}H_{2412}N_{402}O_{471}S_{12}}$} & {\footnotesize -12330} & {\footnotesize -4.7} & {\footnotesize -0.159}\tabularnewline
{\footnotesize TRX3} & {\footnotesize P25372} & {\footnotesize Mitochondrion} & {\footnotesize 127} & {\footnotesize $\mathrm{C_{651}H_{1049}N_{167}O_{181}S_{10}}$} & {\footnotesize -4617} & {\footnotesize 4.9} & {\footnotesize -0.255}\tabularnewline
{\footnotesize TRXB2} & {\footnotesize P38816} & {\footnotesize Mitochondrion} & {\footnotesize 342} & {\footnotesize $\mathrm{C_{1640}H_{2615}N_{449}O_{501}S_{14}}$} & {\footnotesize -12841} & {\footnotesize -1.5} & {\footnotesize -0.145}\tabularnewline
\hline
\end{tabular}
\par\end{centering}

\textbf{a.} Amino acid compositions of subcellular isoforms of glutaredoxin
(GLRX), thioredoxin (TRX) and thioredoxin reductase (TRXB) in \emph{S.
cerevisiae} were taken from the SWISS-PROT database \cite{BBA+03}
(accession numbers shown in the table). Chemical formulas of nonionized
proteins, and calculated standard molal Gibbs energy of formation
from the elements ($\Delta G^{\circ}$, in kcal mol$^{-1}$, at 25
$^{\circ}$C and 1 bar) and net ionization state ($Z$) at $\mathrm{pH=7}$
of charged proteins are listed. Average nominal oxidation state of
carbon ($\overline{Z}_{\mathrm{C}}$) was calculated using Eqn. (\ref{eq:ZC}).
\end{table}

Calculated metastability relations are described below for intercompartmental
interactions between the model homologs and proteologs, and for intracompartmental
interactions among the most abundant proteins in compartments or the
reference model complexes. Experimental comparisons and discussion
of their implications are integrated with these results.

\subsection*{Relative metastabilities of subcellular homologs of redoxins}

The cytoplasmic, nuclear and mitochondrial homologs of glutaredoxin
\cite{PPM+02,MBD+04,HRT+06} and thioredoxin/thioredoxin reductase
\cite{PKM+99,TG05} in yeast cells represent the first model systems
for subcellular environments studied here. The names and chemical
formulas of these proteins are listed in Table \ref{table:homologs},
together with some computed properties. The average nominal oxidation
state of carbon ($\overline{Z}_{\mathrm{C}}$) is a function of the
relative proportions of the elements in the chemical formula (see
Methods). These values are provided just to get some initial bearing
on the differences in compositions of the proteins. In Table \ref{table:homologs}
the proteins with the lowest values of $\overline{Z}_{C}$ are the
mitochondrial homologs and those with the highest values of $\overline{Z}_{C}$
are the nuclear homologs.

\begin{wrapfigure}{O}{0.6\columnwidth}%
\begin{centering}
\includegraphics[width=0.55\columnwidth]{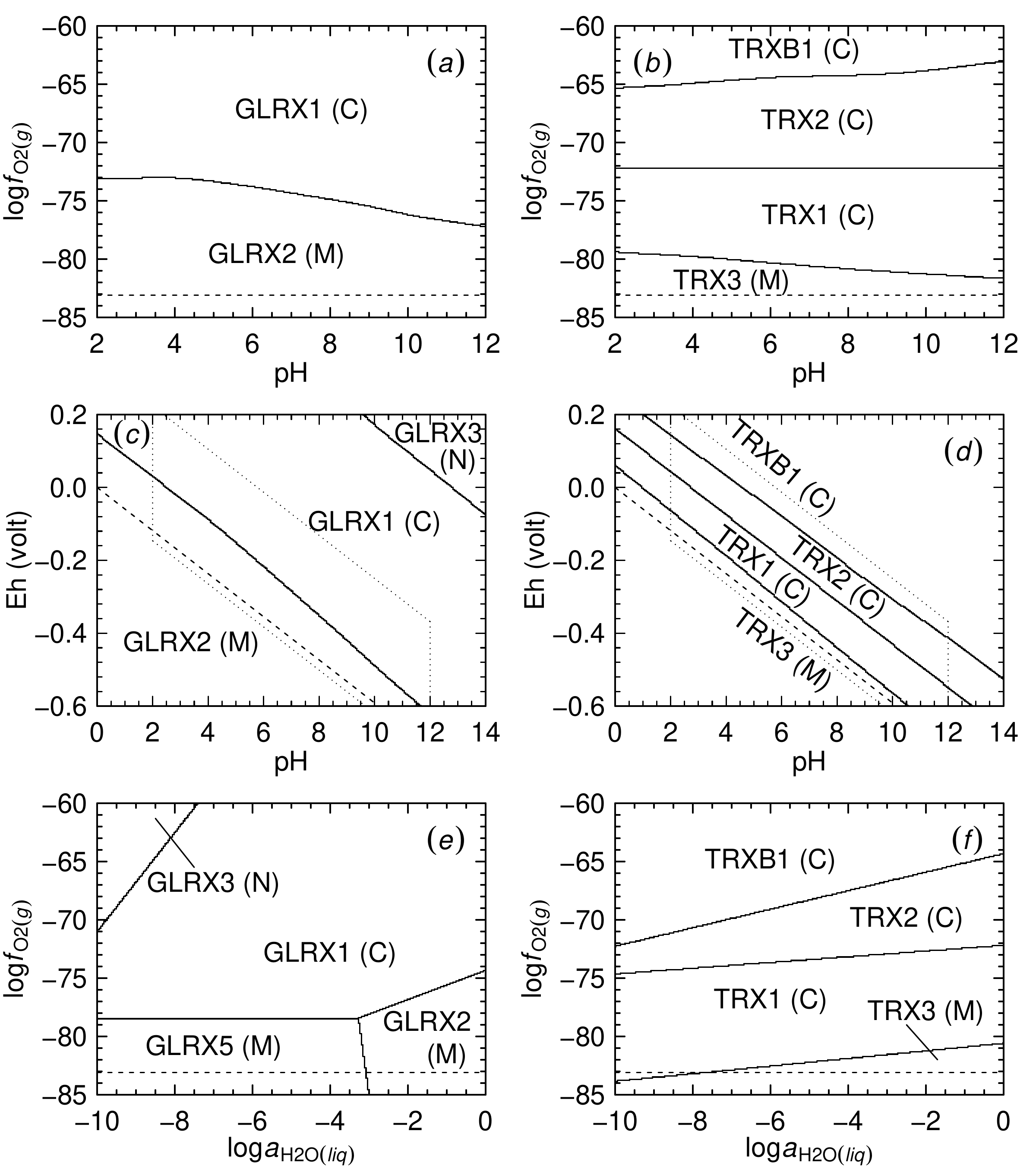}
\par\end{centering}

\caption{\label{fig:redoxin}\textbf{Relative metastabilities of homologs of
glutaredoxin and thioredoxin/thioredoxin reductase.} Predominance
diagrams were generated for homologs of (\emph{a,c},\emph{e}) glutaredoxin
and of (\emph{b},\emph{d},\emph{f}) thioredoxin/thioredoxin reductase
in \emph{S. cerevisiae}. The letters in parentheses following the
labels indicate the subcellular compartment to which the protein is
localized (C -- cytoplasm; M -- mitochondrion; N -- nucleus). Calculations
were performed for ionized proteins at 25 $^{\circ}$C and 1 bar and
for reference activities of basis species noted in the Methods. Reduction
stability limits of $\mathrm{H_{2}O}$ are shown by dashed lines;
the dotted lines in (\emph{c}) and (\emph{d}) correspond to the plot
limits of (\emph{a}) and (\emph{b}).}
\end{wrapfigure}%

Because the current objective is to describe the compositions of populations
of proteins in terms of a variable like oxidation-reduction potential,
a quantity such as $\overline{Z}_{C}$ is not sufficient; it has no
explicitly derivable relation to intensive properties that can be
measured. The forces acting on chemical transformations among proteins
can, however, be assessed by first writing chemical reactions denoting
their formation. An example of this procedure is given further below
for a specific model system. The basic methods that apply there were
used throughout this study. The standard molal Gibbs energies ($\Delta G^{\circ}$)
and net charges of ionized proteins at $\mathrm{pH=7}$ are listed
in Table \ref{table:homologs} so that the results described below
can be reproduced at this $\mathrm{pH}$.

In Figs. \ref{fig:redoxin}a and b the metastable equilibrium predominance
limits of ionized proteins in the glutaredoxin and thioredoxin/thioredoxin
reductase model systems are shown as a function of the logarithm of
oxygen fugacity and $\mathrm{pH}$. Here, the predominant protein
in a population is taken to be the one with the greatest chemical
activity. The computation of the relative metastabilities of the proteins
included all five model proteins in the glutaredoxin system as candidates,
but note regarding Fig. \ref{fig:redoxin}a that only two of the five
proteins appear on the diagram. Those that do not appear are less
metastable, or have greater energy requirements for their formation
over the range of conditions represented in Fig. \ref{fig:redoxin}a
than either of the proteins appearing in the figure. 

The equal-activity lines in these $\mathrm{pH}$ diagrams are curved
because the ionization states of the proteins depend on $\mathrm{pH}$.
The observation apparent in Fig. \ref{fig:redoxin}a that increasing
$\log f_{\mathrm{O_{2}}_{\left(g\right)}}$ favors formation of the
cytoplasmic protein homolog relative to its mitochondrial counterpart
is also true for the thioredoxin/thioredoxin reductase system shown
in Fig. \ref{fig:redoxin}b. In comparing Figs. \ref{fig:redoxin}a
and b note that in the latter figure, predominance fields for a greater
number of candidate proteins appear, and that the predominance field
boundary between mitochondrial and cytoplasmic proteins occurs at
a lower oxidation-reduction potential. The dashed lines shown in each
diagram of Fig. \ref{fig:redoxin} are reference lines denoting the
reduction stability limit of $\mathrm{H_{2}O}$ ($\log f_{\mathrm{O_{2}}_{\left(g\right)}}\approx-83.1$
at 25 $^{\circ}$C and 1 bar \cite{Gar60}).

Predominance diagrams as a function of $\mathrm{Eh}$ and $\mathrm{pH}$
for the glutaredoxin and thioredoxin/thioredoxin reductase systems
are shown in Figs. \ref{fig:redoxin}c and d. Like $\log f_{\mathrm{O_{2}}_{\left(g\right)}}$,
$\mathrm{Eh}$ and $\mathrm{pH}$ together are a measure of the oxidation-reduction
potential of the system; the different scales can be converted using
Eqn. (\ref{eq:Eh-logfO2}). The trapezoidal areas bounded by dotted
lines in Figs. \ref{fig:redoxin}c and d show the ranges of $\mathrm{Eh}$
and $\mathrm{pH}$ corresponding to the $\log f_{\mathrm{O_{2}}_{\left(g\right)}}$-$\mathrm{pH}$
diagrams of Figs. \ref{fig:redoxin}a and b. It can be deduced from
these diagrams that if the upper $\log f_{\mathrm{O_{2}}_{\left(g\right)}}$
limit of Fig. \ref{fig:redoxin}a were extended upward, this diagram
would include a portion of the predominance field for the nuclear
protein GLRX3.

It appears from Figs. \ref{fig:redoxin}a-b that increasing increasing
$\log f_{\mathrm{O_{2}}_{\left(g\right)}}$ at constant $\mathrm{pH}$,
or increasing $\mathrm{pH}$ at constant oxidation-reduction potential
have similar consequences for the relative metastabilities of the
cytoplasmic and mitochondrial homologs. In this analysis, however,
$\mathrm{pH}$ does not appear to be a very descriptive variable;
the magnitude of the effect of changing oxygen fugacity over several
log units is greater than the effect of changing $\mathrm{pH}$ by
several units. In further metastability calculations $\mathrm{pH}$
was set to 7. Also, because $\mathrm{Eh}$ itself is defined in terms
of $\mathrm{pH}$, the oxidation-reduction potential variable adopted
below is $\log f_{\mathrm{O}_{2\left(g\right)}}$, which is more directly
related to the potential of a thermodynamic component.

In Figs. \ref{fig:redoxin}e and f the logarithm of activity of water
($\log a_{\mathrm{H_{2}O}}$) appears as a variable. In Fig. \ref{fig:redoxin}e
it can be seen that the formation of a nuclear homolog of GLRX is
favored relative to the cytoplasmic homologs by decreasing activity
of water and/or increasing oxygen fugacity, and that increasing relative
metastabilities of the mitochondrial proteins are consistent with
lower oxidation-reduction potentials and to some extent higher activities
of water. In Fig. \ref{fig:redoxin}f it appears that the formation
of the thioredoxin reductases relative to thioredoxins in each compartment
is favored by increasing $f_{\mathrm{O_{2}}_{\left(g\right)}}$, and
that for the TRX the relative metastabilities of the mitochondrial
proteins increase with decreasing $f_{\mathrm{O_{2}}_{\left(g\right)}}$.

\subsection*{Comparison with subcellular redox measurements}

\begin{wraptable}{O}{0.65\columnwidth}%

\caption{\label{table:environment}Nominal electrochemical characteristics
of subcellular environments in eukaryotes. Values refer to yeast cells
unless noted otherwise.}

\begin{centering}
\begin{tabular}{lrrr}
\hline 
\multicolumn{1}{l}{{\footnotesize Environment}} & {\footnotesize $\mathrm{Eh}$, volt} & {\footnotesize $\mathrm{pH}$} & \multicolumn{1}{c}{{\footnotesize $\log f_{\mathrm{O_{2}}_{\left(g\right)}}$$^{\mathbf{m}}$}}\tabularnewline
\hline
{\footnotesize Extracellular (intestine)} & {\footnotesize $-0.137$ to $-0.80$$^{\mathbf{a}}$} & {\footnotesize $3$$^{\mathbf{g}}$} & {\footnotesize $-83.3$ to $-79.4$}\tabularnewline
{\footnotesize Cytoplasm} & {\footnotesize $-0.235$ to $-0.222$$^{\mathbf{b}}$} & {\footnotesize $6.5$$^{\mathbf{h}}$} & {\footnotesize $-75.9$ to $-75.0$}\tabularnewline
{\footnotesize Nucleus} & {\footnotesize --$^{\mathbf{c}}$} & {\footnotesize $7.7$$^{\mathbf{i}}$} & {\footnotesize --$^{\mathbf{c}}$}\tabularnewline
{\footnotesize Mitochondrion} & {\footnotesize $-0.360$$^{\mathbf{d}}$} & {\footnotesize $8$$^{\mathbf{j}}$} & {\footnotesize $-78.3$}\tabularnewline
{\footnotesize Endoplasmic reticulum} & {\footnotesize $-0.185$ to $-0.133$$^{\mathbf{e}}$} & {\footnotesize $7.2$$^{\mathbf{k}}$} & {\footnotesize $-69.7$ to $-66.2$}\tabularnewline
{\footnotesize Vacuole} & {\footnotesize $>+0.769$$^{\mathbf{f}}$} & {\footnotesize $6.2$$^{\mathbf{l}}$} & {\footnotesize $>-9.2$}\tabularnewline
\hline
\end{tabular}
\par\end{centering}

\textbf{a. }\cite{DJ00} (\emph{Homo sapiens}). \textbf{b.} The lower
and upper values are taken from \cite{DTG+05} and \cite{TG03b},
respectively. \textbf{c.} The state of the GSSG/GSH couple in the
nucleus is thought to be more reduced than in the cytoplasm \cite{HGJ06};
see text. \textbf{d.} \cite{HAO+04} (\emph{Homo sapiens} HeLa \cite{MSP+99}
cells). \textbf{e.} \cite{HSL92} (\emph{Mus musculus}: mouse hybridoma
cells \cite{CRL1606}). \textbf{f.} Calculated by combining the law
of mass action for $\mathrm{Fe^{+3}}+e^{-}\rightleftharpoons\mathrm{Fe^{+2}}$
(standard molal Gibbs energies taken from \cite{SSW+97}) with $a_{\mathrm{Fe^{+3}}}=a_{\mathrm{Fe^{+2}}}$
(see text). \textbf{g.} \cite{Moj96} (\emph{Homo sapiens}). \textbf{h.}
\cite{IO95} (yeast). \textbf{i.} \cite{CLW06} (organism unspecified).
\textbf{j.} \cite{LMM+98} (HeLa) \textbf{k.} \cite{WLA+00}. \textbf{l.}
\cite{PMJ89}. \textbf{m.} Values of $\mathrm{Eh}$ and $\mathrm{pH}$
listed here were combined with Eqn. (\ref{eq:Eh-logfO2}) at $T=25$
$^{\circ}$C, $P=1$ bar and $a_{\mathrm{H_{2}O}}=1$ to generate
the values of $\log f_{\mathrm{O_{2}}_{\left(g\right)}}$.

\end{wraptable}%

Let us compare the positions of the predominance fields in Fig. \ref{fig:redoxin}
with measured subcellular redox states. The values of $\mathrm{Eh}$
derived from the concentrations of oxidized and reduced glutathione
(GSSG and GSH, respectively) in extra- and subcellular environments
reported in various studies \cite{HSL92,DJ00,HAO+04,TG03b,DTG+05}
were converted to corresponding values of $\log f_{\mathrm{O_{2}}_{\left(g\right)}}$
using Eqn. (\ref{eq:Eh-logfO2}) in the Methods and are listed in
Table \ref{table:environment}. In order to fill in the table as completely
as possible, it was necessary to consider measurements performed on
eukaryotic cells other than those of \emph{S. cerevisiae} (\emph{e.g.},
HeLa \cite{MSP+99} and mouse hybridoma \cite{CRL1606} cells). The
values of $\mathrm{pH}$ required for conversion of $\mathrm{Eh}$
to $\log f_{\mathrm{O_{2}}_{\left(g\right)}}$ were also retrieved
from the literature \cite{Moj96,IO95,LMM+98}. The computation of
$\log f_{\mathrm{O_{2}}_{\left(g\right)}}$ from $\mathrm{Eh}$ was
performed at 25 $^{\circ}$C and 1 bar and with $\log a_{\mathrm{H_{2}O}}=0$.
No measurements of vacuolar $\mathrm{Eh}$ have been reported, but
it has been noted that $\mathrm{Fe^{+3}}$ predominates over $\mathrm{Fe^{+2}}$
in this compartment \cite{SKK07}. Hence, a nominal (and relatively
very oxidizing) value of $\mathrm{Eh}$ for the vacuole was calculated
that corresponds to equal activities of $\mathrm{Fe^{+3}}$ and $\mathrm{Fe^{+2}}$.

The available measurements of redox states in compartments of eukaryotic
cells can be summarized as, from most reducing to most oxidizing,
mitochondria - nucleus - cytoplasm - endoplasmic reticulum - extracellular
\cite{HGJ06}. Strong redox gradients within the mitochondrion are
essential to its function \cite{GJ08}, which is not captured by the
single values listed in Table \ref{table:environment}. Comparison
nevertheless with the computational results shown in Fig. \ref{fig:redoxin}
indicates that a relatively reducing environment does metastably favor
the mitochondrial homolog. 

Measurements of GSH/GSSG concentrations point to a lower redox state
in the nucleus than in the cytoplasm, but the chemical thermodynamic
predictions show the nuclear proteins favored by relatively oxidizing
conditions. Studies using nuclear magnetic resonance (NMR) showing
that the hydration state of the nucleus is higher than the cytoplasm
\cite{PZK+95,MKO+96} bring into question the prediction consistent
with Fig. \ref{fig:redoxin}e that the formation of the nuclear proteins
is favored relative to their cytoplasmic counterparts by decreasing
activity of water.  Also, mitochondrial $\mathrm{pH}$ is somewhat
higher than that of the cytoplasm \cite{IO95,LMM+98}, but in Figs.
\ref{fig:redoxin}a and b it appears that the predicted energetic
constraints favor the cytoplasmic proteins at higher $\mathrm{pH}$s.
These comparisons indicate that all metastable equilibrium constraints
are not preserved in the spatial relationships of the homologous redoxins
in the cell.

\begin{table}
\begin{centering}
\caption{\label{table:proteologs}Overall protein compositions (proteologs)
of compartments in yeast cells$^{\mathbf{a}}$.}

\par\end{centering}

\begin{centering}
\begin{tabular}{lrrlrrrr}
\hline 
{\footnotesize Location} & \multicolumn{1}{c}{{\footnotesize Number}} & \multicolumn{1}{c}{{\footnotesize Length}} & \multicolumn{1}{c}{{\footnotesize Formula}} & {\footnotesize $\Delta G^{\circ}$} & {\footnotesize $Z$} & {\footnotesize $\overline{Z}_{\mathrm{C}}$} & {\footnotesize $\log f_{\mathrm{O_{2}}_{\left(g\right)}}$}\tabularnewline
\hline
{\footnotesize actin} & {\footnotesize 22} & {\footnotesize 469.4} & {\footnotesize $\mathrm{C_{2316.7}H_{3636.4}N_{632.4}O_{721.5}S_{10}}$} & {\footnotesize -18500} & {\footnotesize -5.2} & {\footnotesize -0.119} & {\footnotesize -75.0}\tabularnewline
{\footnotesize ambiguous} & {\footnotesize 123} & {\footnotesize 821.3} & {\footnotesize $\mathrm{C_{4108.2}H_{6372.5}N_{1090.7}O_{1267.6}S_{30.3}}$} & {\footnotesize -32181} & {\footnotesize -22.4} & {\footnotesize -0.123} & {\footnotesize NA}\tabularnewline
{\footnotesize bud} & {\footnotesize 57} & {\footnotesize 429.9} & {\footnotesize $\mathrm{C_{2204.9}H_{3403.4}N_{574.6}O_{631.4}S_{19.7}}$} & {\footnotesize -15314} & {\footnotesize 6.8} & {\footnotesize -0.171} & {\footnotesize -75.2}\tabularnewline
{\footnotesize bud.neck} & {\footnotesize 11} & {\footnotesize 905.2} & {\footnotesize $\mathrm{C_{4543.6}H_{7203.6}N_{1250.2}O_{1443.7}S_{26.6}}$} & {\footnotesize -38092} & {\footnotesize -16.8} & {\footnotesize -0.113} & {\footnotesize -69.2}\tabularnewline
{\footnotesize cell.periphery} & {\footnotesize 38} & {\footnotesize 826.1} & {\footnotesize $\mathrm{C_{4178.9}H_{6506.4}N_{1098.2}O_{1229.9}S_{33.8}}$} & {\footnotesize -30639} & {\footnotesize 1.8} & {\footnotesize -0.164} & {\footnotesize -74.7}\tabularnewline
{\footnotesize cytoplasm} & {\footnotesize 746} & {\footnotesize 458.5} & {\footnotesize $\mathrm{C_{2294.5}H_{3623.3}N_{627.9}O_{704.8}S_{13.9}}$} & {\footnotesize -18106} & {\footnotesize -3.9} & {\footnotesize -0.132} & {\footnotesize -74.6}\tabularnewline
{\footnotesize early.Golgi} & {\footnotesize 9} & {\footnotesize 622.9} & {\footnotesize $\mathrm{C_{3197.7}H_{5067.7}N_{821.1}O_{972.8}S_{21.6}}$} & {\footnotesize -25456} & {\footnotesize -19.9} & {\footnotesize -0.193} & {\footnotesize -75.3}\tabularnewline
{\footnotesize endosome} & {\footnotesize 30} & {\footnotesize 457.6} & {\footnotesize $\mathrm{C_{2309.0}H_{3651.3}N_{626.1}O_{721.0}S_{13.9}}$} & {\footnotesize -18833} & {\footnotesize -9.4} & {\footnotesize -0.131} & {\footnotesize -76.7}\tabularnewline
{\footnotesize ER} & {\footnotesize 197} & {\footnotesize 309.7} & {\footnotesize $\mathrm{C_{1670.8}H_{2566.1}N_{415.5}O_{444}S_{10.9}}$} & {\footnotesize -10292} & {\footnotesize 9.4} & {\footnotesize -0.245} & {\footnotesize -75.5}\tabularnewline
{\footnotesize ER.to.Golgi} & {\footnotesize 5} & {\footnotesize 594.9} & {\footnotesize $\mathrm{C_{2951.2}H_{4600.2}N_{790.8}O_{907.1}S_{18.9}}$} & {\footnotesize -22855} & {\footnotesize -13.3} & {\footnotesize -0.127} & {\footnotesize NA}\tabularnewline
{\footnotesize Golgi} & {\footnotesize 14} & {\footnotesize 478.2} & {\footnotesize $\mathrm{C_{2474.6}H_{3863.3}N_{642.7}O_{727.2}S_{15.1}}$} & {\footnotesize -18229} & {\footnotesize -3.4} & {\footnotesize -0.182} & {\footnotesize -75.3}\tabularnewline
{\footnotesize late.Golgi} & {\footnotesize 29} & {\footnotesize 602.4} & {\footnotesize $\mathrm{C_{3024.6}H_{4773.1}N_{802.9}O_{951.3}S_{17.9}}$} & {\footnotesize -24962} & {\footnotesize -22.6} & {\footnotesize -0.141} & {\footnotesize -75.1}\tabularnewline
{\footnotesize lipid.particle} & {\footnotesize 17} & {\footnotesize 502.1} & {\footnotesize $\mathrm{C_{2574.8}H_{3987.7}N_{673.3}O_{752}S_{17.5}}$} & {\footnotesize -18691} & {\footnotesize -4.0} & {\footnotesize -0.167} & {\footnotesize -75.0}\tabularnewline
{\footnotesize microtubule} & {\footnotesize 10} & {\footnotesize 497.0} & {\footnotesize $\mathrm{C_{2508.9}H_{3968.9}N_{689.9}O_{774.4}S_{17.9}}$} & {\footnotesize -20024} & {\footnotesize -3.9} & {\footnotesize -0.125} & {\footnotesize -75.0}\tabularnewline
{\footnotesize mitochondrion} & {\footnotesize 426} & {\footnotesize 484.9} & {\footnotesize $\mathrm{C_{2446.9}H_{3872.8}N_{669.4}O_{725.3}S_{16.0}}$} & {\footnotesize -18244} & {\footnotesize 6.8} & {\footnotesize -0.156} & {\footnotesize -75.9}\tabularnewline
{\footnotesize nuclear.periphery} & {\footnotesize 46} & {\footnotesize 815.6} & {\footnotesize $\mathrm{C_{4110.6}H_{6516.1}N_{1091.9}O_{1272.1}S_{20.6}}$} & {\footnotesize -33153} & {\footnotesize -11.5} & {\footnotesize -0.159} & {\footnotesize -75.2}\tabularnewline
{\footnotesize nucleolus} & {\footnotesize 60} & {\footnotesize 564.3} & {\footnotesize $\mathrm{C_{2788.9}H_{4430.5}N_{771.5}O_{899.0}S_{13.3}}$} & {\footnotesize -23928} & {\footnotesize -10.9} & {\footnotesize -0.104} & {\footnotesize -75.0}\tabularnewline
{\footnotesize nucleus} & {\footnotesize 453} & {\footnotesize 572.1} & {\footnotesize $\mathrm{C_{2843.6}H_{4542.8}N_{802.2}O_{893.6}S_{20.3}}$} & {\footnotesize -23525} & {\footnotesize -2.3} & {\footnotesize -0.108} & {\footnotesize -71.5}\tabularnewline
{\footnotesize peroxisome} & {\footnotesize 18} & {\footnotesize 422.3} & {\footnotesize $\mathrm{C_{2117.9}H_{3334.8}N_{568.6}O_{642.0}S_{13.5}}$} & {\footnotesize -16397} & {\footnotesize -2.0} & {\footnotesize -0.150} & {\footnotesize -74.8}\tabularnewline
{\footnotesize punctate.composite} & {\footnotesize 61} & {\footnotesize 474.7} & {\footnotesize $\mathrm{C_{2355.7}H_{3719.7}N_{643.7}O_{763.3}S_{10.6}}$} & {\footnotesize -20313} & {\footnotesize -22.6} & {\footnotesize -0.102} & {\footnotesize NA}\tabularnewline
{\footnotesize spindle.pole} & {\footnotesize 30} & {\footnotesize 470.9} & {\footnotesize $\mathrm{C_{2391.1}H_{3818.9}N_{659.1}O_{749.4}S_{14.2}}$} & {\footnotesize -19820} & {\footnotesize -6.3} & {\footnotesize -0.131} & {\footnotesize -78.8}\tabularnewline
{\footnotesize vacuolar.membrane} & {\footnotesize 45} & {\footnotesize 762.2} & {\footnotesize $\mathrm{C_{3813.0}H_{5973.7}N_{1010.5}O_{1154.1}S_{26.2}}$} & {\footnotesize -29221} & {\footnotesize -15.5} & {\footnotesize -0.153} & {\footnotesize -75.2}\tabularnewline
{\footnotesize vacuole} & {\footnotesize 67} & {\footnotesize 511.7} & {\footnotesize $\mathrm{C_{2543.3}H_{3903.1}N_{650.7}O_{800.0}S_{16.8}}$} & {\footnotesize -20239} & {\footnotesize -15.9} & {\footnotesize -0.125} & {\footnotesize -70.6}\tabularnewline
\hline
\end{tabular}
\par\end{centering}

\textbf{a.} Chemical formulas of nonionized proteologs and standard
molal Gibbs energy of formation from the elements ($\Delta G^{\circ}$,
in kcal mol$^{-1}$, at 25 $^{\circ}$C and 1 bar) and net ionization
state ($Z$) at $\mathrm{pH=7}$ of ionized proteologs were calculated
using the overall amino acid compositions given in Table S1. Values
of the nominal oxidation state of carbon ($\overline{Z}_{\mathrm{C}}$)
were calculated using Eqn. (\ref{eq:ZC}). $\log f_{\mathrm{O_{2}}_{\left(g\right)}}$
values for compartments were determined from the metastable equilibrium
limits of subcellular interactions listed in Table \ref{table:interactions}.
\end{table}

\subsection*{Relative metastabilities of proteologs}

\begin{wrapfigure}{O}{0.72\columnwidth}%
\begin{centering}
\includegraphics[width=0.7\columnwidth]{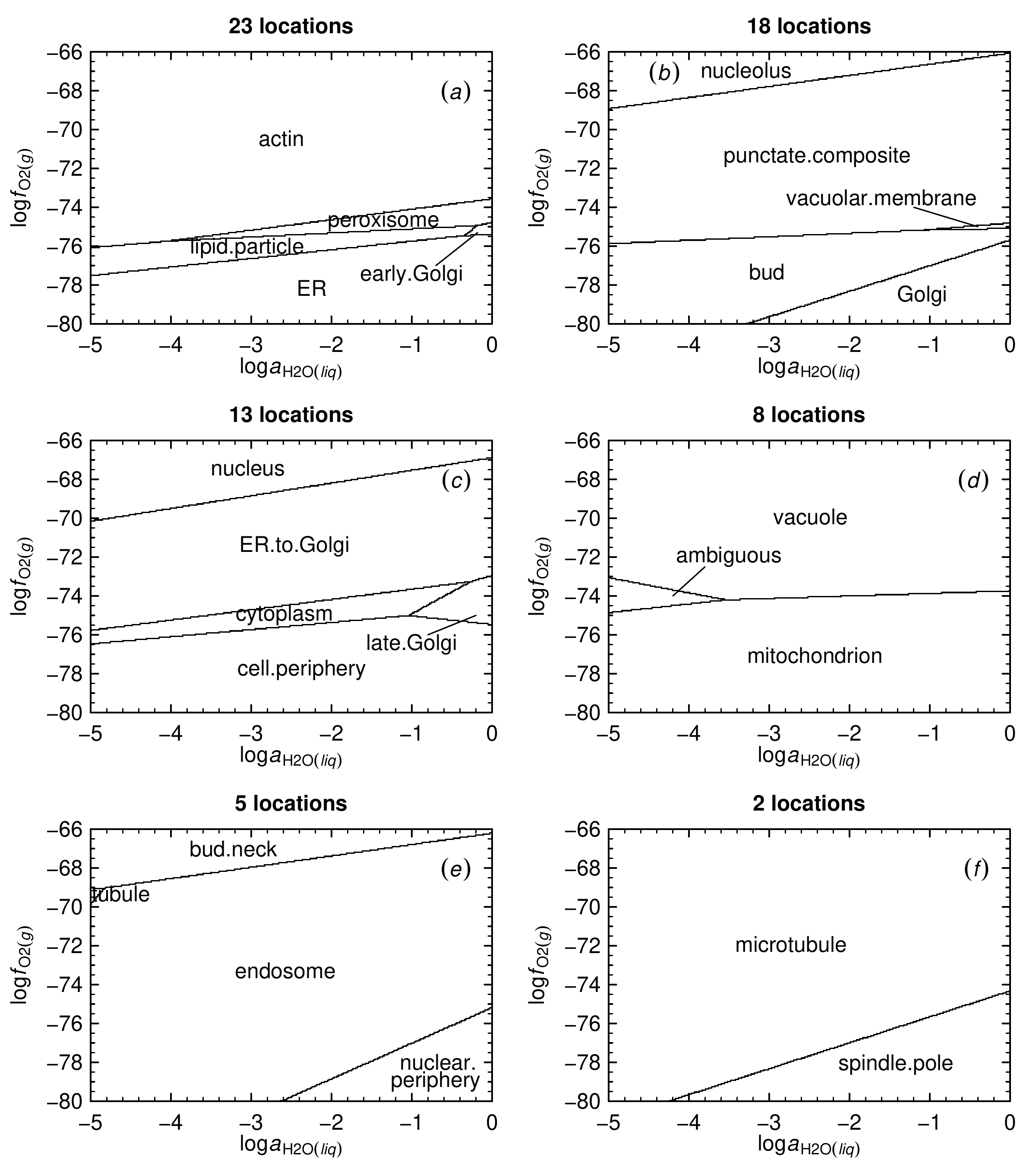}
\par\end{centering}

\caption{\label{fig:proteologs}\textbf{Relative metastabilities of proteologs
of compartments.} Predominance diagrams were generated as a function
of $\log f_{\mathrm{O_{2}}_{\left(g\right)}}$ and $\log a_{\mathrm{H_{2}O}}$
at 25 $^{\circ}$C and 1 bar for the proteologs listed in Table \ref{table:proteologs}.
The diagram in (\emph{a}) represents 23 model proteologs; diagrams
in panels (\emph{b})--(\emph{f}) represent successively fewer model
proteologs.}

\end{wrapfigure}

The chemical formulas and thermodynamic properties of the model proteologs
-- hypothetical proteins representing the overall amino acid compositions
of compartments (see Methods) -- are listed in Table \ref{table:proteologs}.
The predominance diagrams in Fig. \ref{fig:proteologs} depicting
the relative metastabilities of the model proteologs as a function
of $\log f_{\mathrm{O_{2}}_{\left(g\right)}}$ and $\log a_{\mathrm{H_{2}O}}$
were generated in sequential order. The first diagram in this figure
corresponds to a system in which all 23 proteologs were considered.
Subsequent diagrams in Fig. \ref{fig:proteologs} were generated by
eliminating from consideration some or all of the proteologs represented
by predominance fields in the immediately preceding diagram. It can
be seen in Fig. \ref{fig:proteologs}a that consideration of 23 proteologs
resulted in predicted predominance fields for six proteins over the
ranges of $\log f_{\mathrm{O_{2}}_{\left(g\right)}}$ and $\log a_{\mathrm{H_{2}O}}$
shown in the diagram. Subsequent diagrams in the sequence represent
proteologs with lower predicted relative metastabilities, i.e., higher
energy requirements for formation relative to proteologs appearing
earlier in the sequence.

There is a large difference between the relatively oxidized conditions
of the endoplasmic reticulum reported in the literature (see Table
\ref{table:environment}) and the theoretically relatively reduced
environment of the ER proteolog shown in Fig. \ref{fig:proteologs}a.
Also note the average nominal carbon oxidation state of the ER proteolog,
which is the lowest of any in Table \ref{table:proteologs}. A possible
interpretation of these observations is that there is significant
chemical heterogeneity within this compartment and a relatively high
energy demand for the formation of these proteins in the oxidizing
spaces. Nevertheless, the juxtaposition in the ER of very reduced
proteins and high redox potential does permit a possible advantage:
If the redox potential of the compartment were much lower, the proteins
constituting the endoplasmic reticulum would become more favorable
to produce than any other proteins (see below) ultimately localized
to other compartments that are initially produced there. Perhaps in
this way a high redox state could signal the production of cytoplasmic
and secreted proteins and a drop in redox state the production of
biosynthetic enzymes, i.e. the reproduction of the ER itself.

The proteologs appearing in successive diagrams in Fig. \ref{fig:proteologs}
are characterized by increasingly higher predicted energy requirements
for their formation. Hence, the nuclear, cytoplasmic and mitochondrial
proteologs appearing in Fig. \ref{fig:proteologs}c-d are relatively
less metastable compared to those of actin, early Golgi and ER appearing
in Fig. \ref{fig:proteologs}a. It is noteworthy that the proteologs
representing the two cytoskeletal systems in yeast cells, actin and
microtubule, appear at opposite ends of the energy spectrum. This
prediction may be consistent with the observation that actin in different
forms appears to be present at most stages of the cell cycle \cite{BAM+97},
but that the microtubule cytoskeleton grows during anaphase (i.e.,
the stage of the cell cycle characterized by physical separation of
the chromosomes; \cite{ABL+89}) and is degraded during other stages
of the cell cycle \cite{BAM+97,ABL+89}.

The order of appearance of phases throughout a reaction sequence is
determined by the relative stabilities of the phases \cite{Hel70c}.
Examples of the application of this notion in inorganic systems are
the reaction series of metamorphic minerals, paragenetic sequences
of mineralization \cite{Pal29}, Ostwald ripening \cite{Nyv95}, and
weathering reaction paths \cite{SLS94}. Can the relative metastabilities
of proteins provide information about their order of appearance in
the cell cycle?

The outcome of the mitotic cycle in \emph{S. cerevisiae} is the growth
of a new cell in the form of a bud \cite{ABL+89}. Not all structures
in the bud form simultaneously. Instead, it has been observed that
\cite{LWP97} {}``the endoplasmic reticulum, Golgi, mitochondria,
and vacuoles all begin to populate the bud well before anaphase and
that their segregation into the bud does not require microtubules''.
The results in Fig. \ref{fig:proteologs} indicate that the proteolog
for bud is of comparable metastability relative to that of Golgi but
it less metastable than the proteolog of ER. In the absence of energy
input, it follows that there would be a chemical driving force to
form the ER proteins at the expense of any of the bud that may be
present. The appearance in the bud of the less-metastable mitochondrial
proteins suggests that there is a source of energy to the bud that
is nevertheless not sufficient to drive the formation of the proteins
in the microtubule. The formation of these proteins may not be possible
until the products of the mitochondrial reactions and other energy-rich
metabolites have accumulated in the cell.

\subsection*{Intercompartmental protein interactions}

The diagrams in Fig. \ref{fig:proteologs} show the predominant metastability
interactions between proteologs for different subcellular compartments.
However, many subcellular interactions may in fact be meta-metastable
with respect to Fig. \ref{fig:proteologs}. For example, interactions
occur between proteins in the cytoplasm and nucleus \cite{WGC97},
but the proteologs for these compartments do not share a reaction
boundary in Fig. \ref{fig:proteologs}c. Below, known intercompartmental
interactions are combined with the oxygen fugacity requirements for
(meta-)metastable equilibrium of the proteologs to characterize compartmental
oxidation-reduction potentials. These are used in the next section
to explore a possible developmental reaction path.

\begin{table}
\caption{\label{table:interactions}Major intercompartmental protein interactions
in yeast$^{\mathbf{a}}$.}

\begin{centering}
\begin{tabular}{lrrlrr}
\hline 
{\footnotesize Interaction} & {\footnotesize $\Delta n_{\mathrm{O_{2}}}$} & {\footnotesize $\log f_{\mathrm{O_{2}}_{\left(g\right)}}$} & {\footnotesize Interaction} & {\footnotesize $\Delta n_{\mathrm{O_{2}}}$} & {\footnotesize $\log f_{\mathrm{O_{2}}_{\left(g\right)}}$}\tabularnewline
\hline
\textbf{\footnotesize actin}{\footnotesize --}\textbf{\footnotesize bud} & {\footnotesize 0.266} & {\footnotesize -75.1} & \textbf{\footnotesize vacuole}{\footnotesize --}\textbf{\footnotesize bud} & {\footnotesize 0.223} & {\footnotesize -73.8}\tabularnewline
\textbf{\footnotesize actin}{\footnotesize --bud.neck} & {\footnotesize 0.078} & {\footnotesize -83.4} & \textbf{\footnotesize vacuole}{\footnotesize --}\textbf{\footnotesize cell.periphery} & {\footnotesize 0.140} & {\footnotesize -73.4}\tabularnewline
\textbf{\footnotesize actin}{\footnotesize --}\textbf{\footnotesize cell.periphery} & {\footnotesize 0.183} & {\footnotesize -75.4} & \textbf{\footnotesize vacuole}{\footnotesize --}\textbf{\footnotesize cytoplasm} & {\footnotesize 0.044} & {\footnotesize -70.7}\tabularnewline
\textbf{\footnotesize actin}{\footnotesize --}\textbf{\footnotesize endosome} & {\footnotesize 0.129} & {\footnotesize -76.6} & \textbf{\footnotesize vacuole}{\footnotesize --}\textbf{\footnotesize endosome} & {\footnotesize 0.086} & {\footnotesize -74.1}\tabularnewline
\textbf{\footnotesize actin}{\footnotesize --}\textbf{\footnotesize vacuolar.membrane} & {\footnotesize 0.111} & {\footnotesize -75.1} & \textbf{\footnotesize vacuole}{\footnotesize --}\textbf{\footnotesize late.Golgi} & {\footnotesize 0.072} & {\footnotesize -71.5}\tabularnewline
\textbf{\footnotesize actin}{\footnotesize --}\textbf{\footnotesize mitochondrion} & {\footnotesize 0.161} & {\footnotesize -75.8} & {\footnotesize nucleus--}\textbf{\footnotesize actin} & {\footnotesize -0.023} & {\footnotesize -88.7}\tabularnewline
\textbf{\footnotesize actin}{\footnotesize --microtubule} & {\footnotesize 0.124} & {\footnotesize -78.3} & \textbf{\footnotesize nucleus}{\footnotesize --microtubule} & {\footnotesize 0.101} & {\footnotesize -75.9}\tabularnewline
{\footnotesize microtubule--}\textbf{\footnotesize bud} & {\footnotesize 0.142} & {\footnotesize -72.3} & \textbf{\footnotesize nucleus}{\footnotesize --}\textbf{\footnotesize spindle.pole} & {\footnotesize 0.139} & {\footnotesize -75.5}\tabularnewline
\textbf{\footnotesize microtubule}{\footnotesize --}\textbf{\footnotesize bud.neck} & {\footnotesize -0.045} & {\footnotesize -69.4} & \textbf{\footnotesize nucleus}{\footnotesize --}\textbf{\footnotesize bud} & {\footnotesize 0.243} & {\footnotesize -73.8}\tabularnewline
{\footnotesize microtubule--}\textbf{\footnotesize cell.periphery} & {\footnotesize 0.059} & {\footnotesize -69.2} & \textbf{\footnotesize nucleus}{\footnotesize --bud.neck} & {\footnotesize 0.056} & {\footnotesize -81.3}\tabularnewline
{\footnotesize microtubule--}\textbf{\footnotesize cytoplasm} & {\footnotesize -0.037} & {\footnotesize -83.3} & \textbf{\footnotesize nucleus}{\footnotesize --}\textbf{\footnotesize cytoplasm} & {\footnotesize 0.064} & {\footnotesize -71.7}\tabularnewline
{\footnotesize microtubule--}\textbf{\footnotesize spindle.pole} & {\footnotesize 0.038} & {\footnotesize -74.3} & {\footnotesize nucleus--}\textbf{\footnotesize nucleolus} & {\footnotesize -0.034} & {\footnotesize -78.7}\tabularnewline
\textbf{\footnotesize spindle.pole}{\footnotesize --}\textbf{\footnotesize cytoplasm} & {\footnotesize -0.075} & {\footnotesize -78.7} & \textbf{\footnotesize nuclear.periphery}{\footnotesize --}\textbf{\footnotesize bud.neck} & {\footnotesize -0.080} & {\footnotesize -69.3}\tabularnewline
{\footnotesize spindle.pole--}\textbf{\footnotesize nuclear.periphery} & {\footnotesize -0.004} & {\footnotesize -119.1} & {\footnotesize nuclear.periphery--}\textbf{\footnotesize cytoplasm} & {\footnotesize -0.072} & {\footnotesize -76.5}\tabularnewline
\textbf{\footnotesize ER}{\footnotesize --}\textbf{\footnotesize cell.periphery} & {\footnotesize -0.460} & {\footnotesize -74.7} & \textbf{\footnotesize nuclear.periphery}{\footnotesize --}\textbf{\footnotesize nucleus} & {\footnotesize -0.136} & {\footnotesize -74.2}\tabularnewline
\textbf{\footnotesize ER}{\footnotesize --}\textbf{\footnotesize cytoplasm} & {\footnotesize -0.557} & {\footnotesize -74.7} & \textbf{\footnotesize nuclear.periphery}{\footnotesize --}\textbf{\footnotesize nucleolus} & {\footnotesize -0.169} & {\footnotesize -75.1}\tabularnewline
\textbf{\footnotesize ER}{\footnotesize --}\textbf{\footnotesize early.Golgi} & {\footnotesize -0.345} & {\footnotesize -75.4} & \textbf{\footnotesize peroxisome}{\footnotesize --cell.periphery} & {\footnotesize 0.062} & {\footnotesize -78.7}\tabularnewline
\textbf{\footnotesize ER}{\footnotesize --nuclear.periphery} & {\footnotesize -0.485} & {\footnotesize -74.4} & \textbf{\footnotesize peroxisome}{\footnotesize --cytoplasm} & {\footnotesize -0.034} & {\footnotesize -67.2}\tabularnewline
\textbf{\footnotesize ER}{\footnotesize --}\textbf{\footnotesize peroxisome} & {\footnotesize -0.522} & {\footnotesize -75.2} & \textbf{\footnotesize peroxisome}{\footnotesize --}\textbf{\footnotesize lipid.particle} & {\footnotesize 0.138} & {\footnotesize -74.9}\tabularnewline
\textbf{\footnotesize Golgi}{\footnotesize --endosome} & {\footnotesize -0.199} & {\footnotesize -74.3} & \textbf{\footnotesize peroxisome}{\footnotesize --mitochondrion} & {\footnotesize 0.040} & {\footnotesize -82.6}\tabularnewline
\textbf{\footnotesize Golgi}{\footnotesize --}\textbf{\footnotesize vacuole} & {\footnotesize -0.285} & {\footnotesize -74.2} & {\footnotesize mitochondrion--}\textbf{\footnotesize cell.periphery} & {\footnotesize 0.023} & {\footnotesize -72.0}\tabularnewline
\textbf{\footnotesize Golgi}{\footnotesize --}\textbf{\footnotesize late.Golgi} & {\footnotesize -0.213} & {\footnotesize -75.2} & \textbf{\footnotesize mitochondrion}{\footnotesize --}\textbf{\footnotesize cytoplasm} & {\footnotesize -0.074} & {\footnotesize -75.4}\tabularnewline
{\footnotesize Golgi--}\textbf{\footnotesize early.Golgi} & {\footnotesize -0.030} & {\footnotesize -84.4} & \textbf{\footnotesize mitochondrion}{\footnotesize --}\textbf{\footnotesize nucleus} & {\footnotesize -0.138} & {\footnotesize -73.7}\tabularnewline
\hline
\end{tabular}
\par\end{centering}

\textbf{a.} Interactions between proteins in different subcellular
locations in \emph{S. cerevisiae} were identified in the literature.
The calculated reaction coefficients on $\mathrm{O_{2}}_{\left(g\right)}$
and the metastable equilibrium value of $\log f_{\mathrm{O_{2}}_{\left(g\right)}}$
were calculated for each reaction between model proteologs. Names
of locations shown in bold indicate that the model value of $\log f_{\mathrm{O_{2}}_{\left(g\right)}}$
for this compartment (Table \ref{table:proteologs}) lies in the metastability
range for the proteolog in the particular reaction.
\end{table}

To assess the biochemical evidence for specific interactions between
proteins in different compartments in yeast cells, a series of review
papers was surveyed \cite{BAM+97,KGS97,JWH97,WGC97,LK97,PS91}. Statements
implying interaction between proteins in different compartments were
identified by scanning for action words including interact, are at,
align, end at, organize, embed, move, associate, found, locate, extend,
bisect, move, migrate, enter, attach, translocate, carry, sort, composed
of, line, dock and fuse, recycle, transport, pinch, proceed, reach,
degrade in, deliver, colocalize, contain, associate, separate, protrude,
penetrate, cooperate, crosstalk, anchor, reside, continuous with,
shuttle, oxidize, essential to, convey, arrange, import, and transcribe.
The source statements are listed in Text S3 and simplified pairwise
representations of the interactions are summarized in Table \ref{table:interactions}.
Of 190 possible combinations between any two of the 20 subcellular
compartments (this count excludes the ambiguous location and ER to
Golgi and punctate composite, which did not appear in the literature
survey), 46 interactions were identified through this survey.

Chemical reactions corresponding to each of the interactions listed
in Table \ref{table:interactions} were written between residue equivalents
of the proteologs, with the reactant proteolog being the one on the
left-hand side of the interaction and the product proteolog the one
on the right-hand side. The reactions are listed in Table S2. Corresponding
values of $\Delta n_{\mathrm{O_{2}}_{\left(g\right)}}$ (reaction
coefficient on $\mathrm{O_{2}}_{\left(g\right)}$) are listed in Table
\ref{table:interactions} together with the values of $\log f_{\mathrm{O_{2}}_{\left(g\right)}}$
where the calculated chemical activities of the two proteologs in
each reaction are equal. Note that there are some reactions where
the absolute value of $\Delta n_{\mathrm{O_{2}}_{\left(g\right)}}$
is substantially smaller than the others; these include spindle pole--nuclear
periphery, Golgi--early Golgi and nucleus--actin. Because of the small
value of $\Delta n_{\mathrm{O_{2}}_{\left(g\right)}}$ in these reactions,
the values of $\log f_{\mathrm{O_{2}}_{\left(g\right)}}$ for equal
activities of these proteins tend to be more extreme than for other
reactions. Note that the sign of $\Delta n_{\mathrm{O_{2}}_{\left(g\right)}}$
denotes the thermodynamically favored direction of the reaction as
$\log f_{\mathrm{O_{2}}_{\left(g\right)}}$ is changed from its equal-activity
value; for example, at $\log f_{\mathrm{O_{2}}_{\left(g\right)}}=-75.1$,
the proteologs of actin and bud metastably coexist with equal chemical
activities, but at higher values that of actin predominates in metastable
equilibrium.

\begin{figure}
\begin{centering}
\includegraphics[width=0.8\columnwidth]{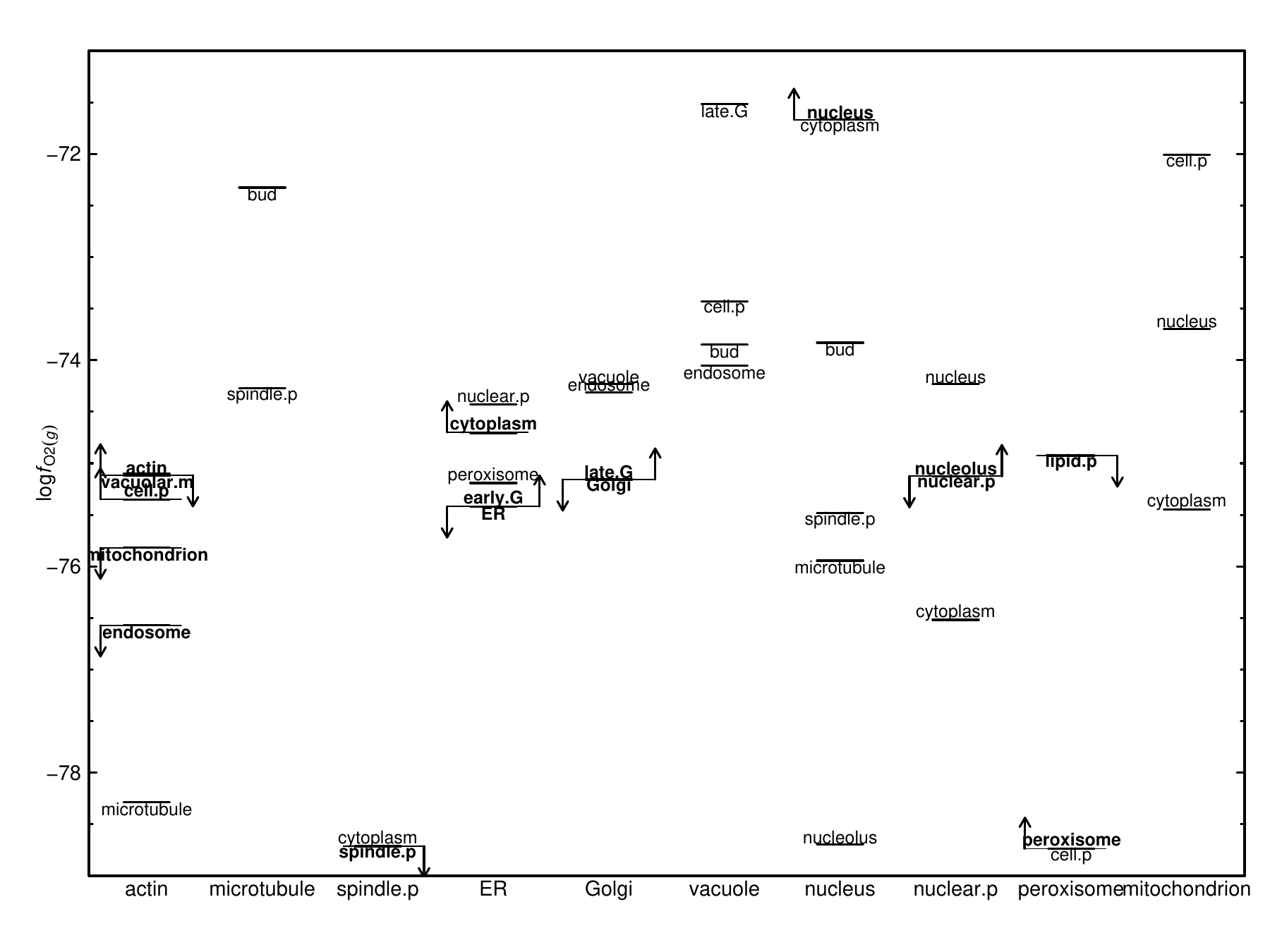}
\par\end{centering}

\caption{\label{fig:interactions}\textbf{Logarithms of oxygen fugacity for
equal chemical activities of proteologs in intercompartmental interactions}.
Metastable equilibrium values of $\log f_{\mathrm{O_{2}}_{\left(g\right)}}$
were obtained for the model reactions listed in Table \ref{table:interactions}.
Reactions are grouped by a common proteolog, listed along the bottom
of the plot. Reactions that were used to derive model values of oxygen
fugacity of compartments listed in Table \ref{table:proteologs} are
denoted by arrows and bold lines and labels. The position of the reaction
labels denotes the direction of the reaction that favors formation
of the corresponding proteolog. The actin--bud and ER--cell periphery
interactions were omitted from this plot to aid in clarity of labeling;
they overlap with actin--vacuolar membrane and ER--cytoplasm, respectively.}

\end{figure}

The interactions listed in Table \ref{table:interactions} were used
to generate model values of the oxygen fugacity in each compartment
that are listed in Table \ref{table:proteologs}. The criterion used
for this analysis was that the oxygen fugacity in a compartment should
in as many cases as possible favor the formation of its proteolog
relative to those of interacting compartments. For example, consider
the proteolog for endosome, which occurs in three interactions listed
in Table \ref{table:interactions}. The endosomal proteolog is favored
to form relative to that of actin by $\log f_{\mathrm{O_{2}}_{\left(g\right)}}<-76.6$
and relative to that of vacuole by $\log f_{\mathrm{O_{2}}_{\left(g\right)}}<-74.1$.
In contrast, the endosomal proteolog is favored to form relative to
the proteolog of Golgi by $\log f_{\mathrm{O_{2}}_{\left(g\right)}}>-74.3$.
A single value of $\log f_{\mathrm{O_{2}}_{\left(g\right)}}$ can
satisfy at most two of these constraints; the model value for endosome
is taken to be just below the limit for its interaction with actin,
or $\log f_{\mathrm{O_{2}}_{\left(g\right)}}=-76.7$ (Table \ref{table:proteologs}).
Because this value favors formation of the endosomal proteolog relative
to those of actin and vacuole, the proteolog of endosome is listed
in bold font in these interactions in Table \ref{table:interactions},
but is shown in normal font in the interaction with the Golgi proteolog.
Similar reasoning was used to derive oxygen fugacities for the other
subcellular compartments listed in Table \ref{table:proteologs},
except for microtubule.

The outcome of the above analysis is summarized in Fig. \ref{fig:proteologs},
where the values of $\log f_{\mathrm{O_{2}}_{\left(g\right)}}$ for
interactions that fall between $-79$ and $-71$ are plotted. The
interactions are grouped by a common interacting proteolog so that
differences between them can be more easily visualized. To avoid clutter,
the reaction labels are generally restricted to the name of a single
proteolog to indicate the direction of $\log f_{\mathrm{O_{2}}_{\left(g\right)}}$
change that favors its formation in the reaction. Model interactions
that were used to constrain the limits of oxygen fugacities for one
compartment (such as the actin--endosome interaction noted above)
or two compartments (such as Golgi--late Golgi) are identified with
one or two arrows, respectively, and the names of the corresponding
proteologs are shown in bold font.

If the model compartmental values of $\log f_{\mathrm{O_{2}}_{\left(g\right)}}$
all favored formation of the corresponding proteologs relative to
their interacting partners, the name of every proteolog would appear
in bold font in Table \ref{table:interactions}. This is only the
case, however, for some proteologs such as that of actin, where $\log f_{\mathrm{O_{2}}_{\left(g\right)}}-75$
favors formation of this proteolog relative to any of its interacting
partners. At the same oxygen fugacity, it can be shown that the proteolog
for microtubule is unmetastable with respect to any of its interacting
partners except for bud neck. Notably, the proteolog for microtubule
only becomes relatively metastable at high oxygen fugacities (w.r.t.
bud, cell periphery and spindle pole) or at low oxygen fugacities
(w.r.t. actin, cytoplasm and nucleus). Hence, the value of $\log f_{\mathrm{O_{2}}_{\left(g\right)}}-75$
taken here for the microtubule compartment is different from all the
others, in that this represents conditions where the formation of
its proteolog is more unfavorable than that of any of its interacting
partners.

\subsection*{Sequential formation driven by oxygen fugacity gradients}

We have already seen theoretical evidence that the microtubule is
a relatively unmetastable assemblage of proteins in the cell. It is
known in spite of this that the microtubule as well as the spindle
pole are essential in cellular division \cite{BAM+97}. Can the metastable
equilibrium relationships reveal anything about the origins of the
interactions of the microtubule and spindle pole in this process?
The following thought experiment explores why the irreversible formation
of proteologs might follow a sequence that is related to metastable
equilibrium thermodynamic relationships.

To start, consider a permeable sac consisting of the cytoplasmic proteolog,
which we will expose to a changing oxidation-reduction environment.
The oxidation-reduction program will begin at $\log f_{\mathrm{O_{2}}_{\left(g\right)}}=-75$,
drop to $\log f_{\mathrm{O_{2}}_{\left(g\right)}}=-83.5$, increase
to $\log f_{\mathrm{O_{2}}_{\left(g\right)}}=-69$ and return to $\log f_{\mathrm{O_{2}}_{\left(g\right)}}=-75$.
At any point along this program the only reactions we will consider
are those involving the proteologs of microtubule or spindle pole.
Let us assume in addition that none of these reactions proceeds to
completion, and that any reaction may only proceed while $\log f_{\mathrm{O_{2}}_{\left(g\right)}}$
is near the equal-activity value for the reaction. Keeping in mind
that no mechanism for the reactions is implied here, it may still
be worthwhile to note that others have observed near-equilibrium concentrations
of substrates in a subset of enzymatically catalyzed reactions \cite{WED74,Vee78}.

At $\log f_{\mathrm{O_{2}}_{\left(g\right)}}=-75$, no reaction occurs
because the conditions coincide with the metastability field of the
cytoplasmic proteolog relative to either microtubule or spindle pole.
As soon as the $\log f_{\mathrm{O_{2}}_{\left(g\right)}}$ decreases
below $-78.7$, some of the spindle pole proteolog may form irreversibly
at the expense of the cytoplasmic proteolog. Below $\log f_{\mathrm{O_{2}}_{\left(g\right)}}=-83.3$,
the microtubular proteolog can begin to form at the expense of the
cytoplasmic proteolog. At $\log f_{\mathrm{O_{2}}_{\left(g\right)}}=-83.5$
both of these reactions may favorably proceed, and we begin now to
increase $\log f_{\mathrm{O_{2}}_{\left(g\right)}}$. As we pass $\log f_{\mathrm{O_{2}}_{\left(g\right)}}=-83.3$,
then $\log f_{\mathrm{O_{2}}_{\left(g\right)}}=-78.7$ going in the
positive direction, some of the proteolog of microtubule, then spindle
pole can react irreversibly to form the cytoplasmic proteolog. These
are the opposite of the first two irreversible reactions.

As long as the current and following reactions do not proceed to completion,
there will be a population of the microtubule and spindle pole proteologs
available to react. Above $\log f_{\mathrm{O_{2}}_{\left(g\right)}}=-78.7$,
where the formation of the cytoplasmic proteolog becomes favored relative
to spindle pole (see above), the proteolog of actin may also favorably
form at the expense of that of microtubule. The nuclear proteolog
can form above $\log f_{\mathrm{O_{2}}_{\left(g\right)}}=-75.9$ at
the expense of the microtubular proteolog, and above $\log f_{\mathrm{O_{2}}_{\left(g\right)}}=-75.5$
at the expense of the spindle pole proteolog. We now momentarily pass
through our starting point, $\log f_{\mathrm{O_{2}}_{\left(g\right)}}=-75$.
So far, the proteologs from spindle pole, microtubule, actin and nucleus,
in that order, may have formed as a result of irreversible reactions
of the original cytoplasmic proteolog. Also, the proteologs of microtubule
and spindle pole may have been subsequently partially degraded after
their possible formation.

Now, as $\log f_{\mathrm{O_{2}}_{\left(g\right)}}$ is increased above
$-74.3$, the proteolog of spindle pole becomes unmetastable relative
to that of microtubule. Above $\log f_{\mathrm{O_{2}}_{\left(g\right)}}=-69.4$,
the proteolog of bud neck may be formed irreversibly at the expense
of that of microtubule. At our maximum $\log f_{\mathrm{O_{2}}_{\left(g\right)}}=-69$
this reaction can continue, but as we drop below $\log f_{\mathrm{O_{2}}_{\left(g\right)}}=-69.2$
it may be joined by formation of the proteolog of cell periphery.
Below $\log f_{\mathrm{O_{2}}_{\left(g\right)}}=-69.4$ any proteolog
of bud neck that may have formed becomes unmetastable relative to
that of microtubule. Below $\log f_{\mathrm{O_{2}}_{\left(g\right)}}=-72.3$
any proteolog of microtubule that remains may degrade in favor of
formation of the proteolog of bud. Finally, as we drop past $\log f_{\mathrm{O_{2}}_{\left(g\right)}}=-74.3$
and return to our starting point of $\log f_{\mathrm{O_{2}}_{\left(g\right)}}=-75$
the proteolog of spindle pole once again becomes relatively metastable
instead of microtubule. In summary, at $\log f_{\mathrm{O_{2}}_{\left(g\right)}}>-75$
the potential arises for formation of proteologs of the microtubule,
bud neck, cell periphery, bud and spindle pole, as well as for retrograde
reactions that may destroy the proteolog of microtubule.

It is important to emphasize the qualified nature of these predictions;
all we know from thermodynamics is that any of these reactions could
have progressed in the direction of a local Gibbs energy minimum.
Whether and to what extent they actually move forward is a consequence
of the reaction mechanism. The purpose of this analysis is not to
suggest any mechanism but to ask whether work performed by control
of $\log f_{\mathrm{O_{2}}_{\left(g\right)}}$ may energize such a
mechanism. The enzymatic properties of the proteins themselves are
probably essential in any actual mechanism. It is encouraging to observe
that at and below the starting $\log f_{\mathrm{O_{2}}_{\left(g\right)}}=-75$
the proteolog of endoplasmic reticulum is favored to form relative
to the cytoplasmic proteolog. Hence under these conditions there exists
a potential for production of biosynthetic enzymes.

\begin{table}
\caption{\label{table:cycle}Hypothetical oxygen fugacity cycle and sequence
of reactions of proteologs.}

\begin{centering}
\begin{tabular}{cccc}
\hline 
$\log f_{\mathrm{O_{2}}_{\left(g\right)}}$ & Reaction & $\log f_{\mathrm{O_{2}}_{\left(g\right)}}$ & Reaction\tabularnewline
\hline
-75.0 & \textbf{Begin} & -74.3 & spindle.pole$\rightarrow$microtubule\tabularnewline
-78.7 & cytoplasm$\rightarrow$spindle.pole & -69.4 & microtubule$\rightarrow$bud.neck\tabularnewline
-83.3 & cytoplasm$\rightarrow$microtubule & -69.0 & \textbf{Maximum point}\tabularnewline
-83.5 & \textbf{Minimum point} & -69.2 & microtubule$\rightarrow$cell.periphery\tabularnewline
-83.3 & microtubule$\rightarrow$cytoplasm & -69.4 & bud.neck$\rightarrow$microtubule\tabularnewline
-78.7 & spindle.pole$\rightarrow$cytoplasm & -72.3 & microtubule$\rightarrow$bud\tabularnewline
-78.7 & microtubule$\rightarrow$actin & -74.3 & microtubule$\rightarrow$spindle.pole\tabularnewline
-75.9 & microtubule$\rightarrow$nucleus & -75.0 & \textbf{End}\tabularnewline
-75.5 & spindle.pole$\rightarrow$nucleus &  & \tabularnewline
\hline
\end{tabular}
\par\end{centering}

\end{table}

The results of this thought experiment are summarized in Table \ref{table:cycle}.
The range of theoretical values of $f_{\mathrm{O_{2}}_{\left(g\right)}}$
required for the chemical transformations among the proteologs is
between $-83.5$ and $-69$, which in terms of redox potential at
25 $^{\circ}$C, 1 bar, $\mathrm{pH=7}$ and $\log a_{\mathrm{H_{2}O}}=0$
correspond to $\mathrm{Eh}=-0.420$V and $\mathrm{Eh=-0.205}$V, respectively
(Eqn. \ref{eq:Eh-logfO2}). The former value is just below the stability
limit for water ($\log f_{\mathrm{O_{2\left(g\right)}}}=-83.1$) but
the redox state of the NADPH/NADP$^{+}$ pool in rat liver mitochondria
might approach this value ($\mathrm{Eh}=-0.415$V \cite{GJ08}). The
latter value is consistent with the state of human cells during differentiation
($\mathrm{Eh}=-0.200$V), which is about $0.040$V higher than proliferating
cells \cite{SB01}.

Oscillations in the redox state of yeast cells are coupled to many
metabolic changes including protein transcription and turnover \cite{LM06}.
Reductive and oxidative phases in the metabolic cycle of yeast have
been identified, with DNA replication occurring during the former
and cell cycle initiation occurring at an advanced stage of the latter
\cite{MG07}. Oxidative stress was shown to hasten HeLa cells into
anaphase by overcoming the normal spindle checkpoint mechanism \cite{DSG07}.
Although the results shown in Table \ref{table:cycle} do not directly
address the synthesis of DNA, they do show that there is a potential
for the formation of the nuclear proteolog during a relatively reducing
part of the hypothetical $f_{\mathrm{O_{2}}_{\left(g\right)}}$ cycle.
In the oxidizing part of this cycle, above $\log f_{\mathrm{O_{2}}_{\left(g\right)}}=-74.3$,
the metastability of the proteolog for spindle pole is decreased,
and at the highest oxidation-reduction potentials a favorable chemical
potential field exists for metastable formation of the proteolog for
bud neck. Hence, the notion that {}``a fundamental redox attractor
underpins ... core cellular processes'' \cite{Mur04} is in principle
supported by the changing relative metastabilities of the proteologs
as a function of oxidation-reduction potential.

\subsection*{Calculation of relative abundances of proteins}

Above, the interactions between homologs (enzyme isoforms) in subcellular
compartments and proteologs representing overall protein compositions
in subcellular compartments were used to derive oxygen fugacity limits
for metastable reaction of proteins in different compartments. In
the second part of this study, attention is focused on the relative
abundances and intracompartmental interactions of proteins.

\begin{figure}
\begin{centering}
\begin{tabular}{cc}
\includegraphics[bb=0bp 432bp 576bp 864bp,clip,width=0.45\columnwidth]{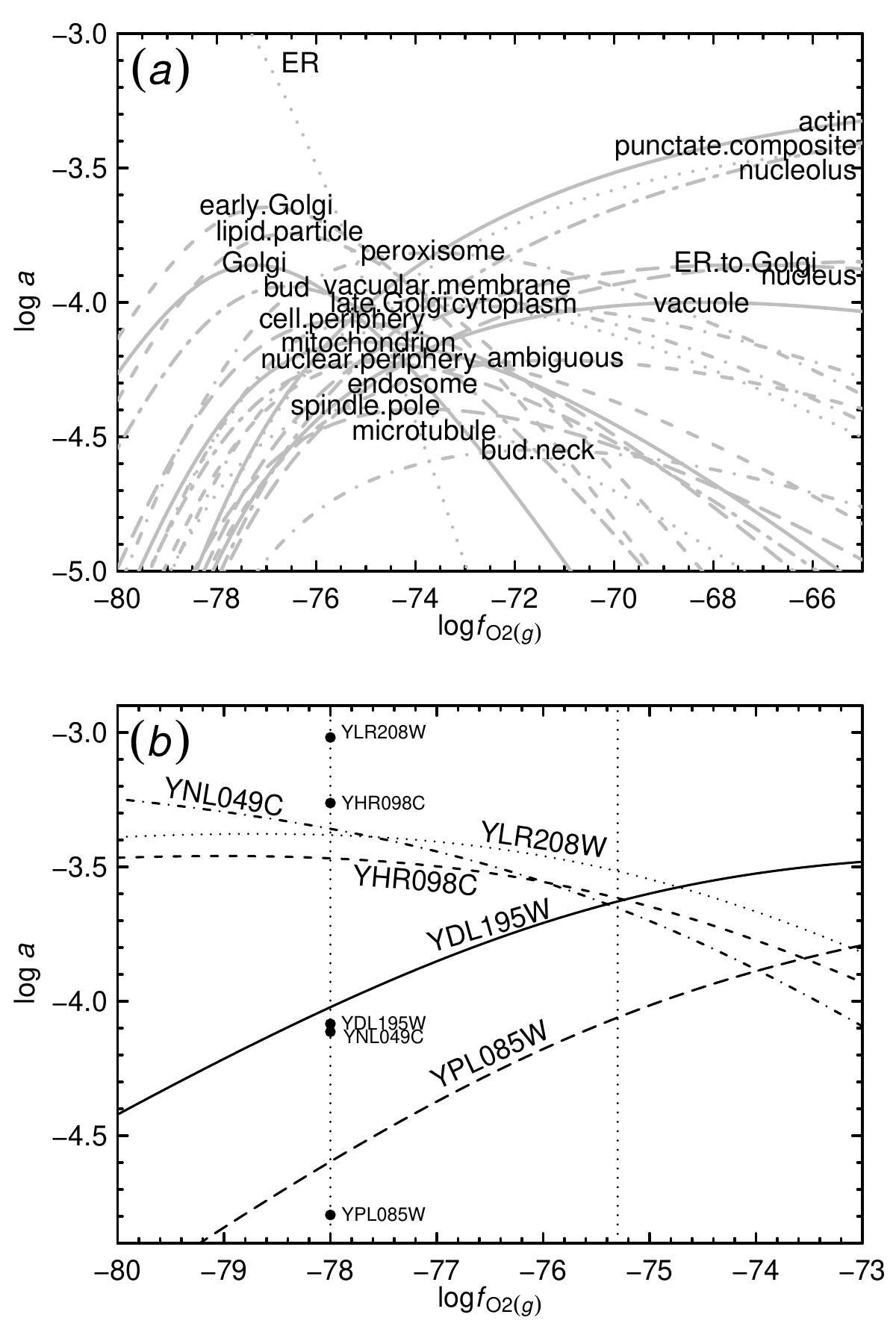} & \includegraphics[bb=0bp 0bp 576bp 432bp,clip,width=0.45\columnwidth]{fig4.pdf}\tabularnewline
\end{tabular}
\par\end{centering}

\caption{\label{fig:loga-logfO2}\textbf{Metastable equilibrium abundances
of model proteologs and proteins as a function of oxygen fugacity.}
Chemical speciation diagrams were generated as a function of $\log f_{\mathrm{O_{2}}_{\left(g\right)}}$
at 25 $^{\circ}$C and 1 bar and with total activity of protein residues
equal to unity for (\emph{a}) the proteologs shown in Table \ref{table:homologs}
and (\emph{b}) the five proteins localized to ER to Golgi whose experimental
abundances were reported in \cite{GHB+03}. The rightmost dotted line
in (\emph{b}) indicates conditions where the calculated abundance
ranking of the proteins is identical to that found in the experiments,
and the leftmost dotted line where the calculated logarithms of activities
have a lower overall deviation from experimental ones, which are indicated
by the points. This value of $\log f_{\mathrm{O_{2}}_{\left(g\right)}}$
($-78$) was used to construct the corresponding diagram in Fig. \ref{fig:proteome_loga}.}

\end{figure}

The logarithms of activities of proteologs consistent with metastable
equilibrium among all 23 model proteologs are plotted in Fig. \ref{fig:loga-logfO2}a
as a function of $\log f_{\mathrm{O_{2}}_{\left(g\right)}}$. This
diagram was generated based on metastable equilibrium among the residues
of the proteins \cite{Dic08a} in the same manner as described in
detail below for a smaller set of proteins (those appearing in Fig.
\ref{fig:loga-logfO2}b). The purpose of Fig. \ref{fig:loga-logfO2}a
is to recapitulate the relationships shown in Fig. \ref{fig:proteologs}.
Note that the same proteins predominate at the extremes of oxygen
fugacity represented in \ref{fig:loga-logfO2}a and in Fig. \ref{fig:redoxin}a
(reducing -- ER; oxidizing -- actin) and that the proteolog of microtubule
appears with low relative abundance. More importantly, perhaps, there
is a minimum in the range of calculated activities of the proteologs
around $\log f_{\mathrm{O_{2}}_{\left(g\right)}}=-75$; changing oxidation-reduction
potential alters not only the identity of the predominant protein
in a metastably interacting population but also the relative abundances
of all the others. There is probably not a single value of $\log f_{\mathrm{O_{2}}_{\left(g\right)}}$
where the calculated relative abundances of the proteologs shown in
Fig. \ref{fig:proteologs} reflect the composition of the cell. Let
us therefore look more closely at the relative abundances of proteins
within compartments.

In Fig. \ref{fig:loga-logfO2}b the relative abundances of the five
model proteins localized exclusively to ER to Golgi are shown as a
function of $\log f_{\mathrm{O_{2}}_{\left(g\right)}}$. A worked-out
example of the calculations leading to this figure, which method also
underlies the generation of the other figures shown here, is presented
in the following paragraphs. 

The model proteins for ER to Golgi, in order of decreasing abundance
in the cell reported by \cite{GHB+03}, are YLR208W, YHR098C, YDL195W,
YNL049C and YPL085W. (For simplicity, the proteins are identified
here by the names of the open reading frames (ORF).) The formula of
the uncharged form of the first protein, YLR208W, is $\mathrm{C}_{1485}\mathrm{H}_{2274}\mathrm{N}_{400}\mathrm{O}_{449}\mathrm{S}_{4}$,
and its amino acid sequence length is 297 residues. The standard molal
Gibbs energy of formation from the elements ($\Delta G^{\circ}$)
of this protein at 25 $^{\circ}$C and 1 bar calculated using group
additivity \cite{DLH06} is $-10670$ kcal mol$^{-1}$ . At this temperature
and pressure and at $\mathrm{pH}=7$, group additivity can also be
used \cite{DLH06} to calculate the charge of the protein ($-10.8832$)
and the standard molal Gibbs energy of formation from the elements
of the charged protein ($-10880$ kcal mol$^{-1}$). The formula of
the protein in this ionization state is $\mathrm{C}_{1485}\mathrm{H}_{2263.1168}\mathrm{N}_{400}\mathrm{O}_{449}\mathrm{S}_{4}^{-10.8832}$.
Dividing by the length of the protein, we find that the formula and
standard molal Gibbs energy of formation from the elements of the
residue equivalent of YLR208W are $\mathrm{C}_{5.0000}\mathrm{H}_{7.6199}\mathrm{N}_{1.3468}\mathrm{O}_{1.5118}\mathrm{S}_{0.0135}^{-0.0366}$
and $-36.633$ kcal mol$^{-1}$, respectively.

The formation from basis species of the residue equivalent of YLR208W
is consistent with\begin{multline}
5.0000\mathrm{CO_{2}}_{\left(aq\right)}+1.7946\mathrm{H_{2}O}+1.3468\mathrm{NH_{3}}_{\left(aq\right)}+0.0135\mathrm{H_{2}S}_{\left(aq\right)}\\
\rightleftharpoons\mathrm{C}_{5.0000}\mathrm{H}_{7.6199}\mathrm{N}_{1.3468}\mathrm{O}_{1.5118}\mathrm{S}_{0.0135}^{-0.0366}+5.1414\mathrm{O_{2}}_{\left(g\right)}+0.0366\mathrm{H^{+}}\,.\label{react:YLR208W}\end{multline}
Similar reasoning can be applied to write the formation reaction of
the residue equivalent of YHR098C as\begin{multline}
4.9720\mathrm{CO}_{2\left(aq\right)}+1.8708\mathrm{H_{2}O}+1.3240\mathrm{NH_{3}}_{\left(aq\right)}+0.0441\mathrm{H_{2}S}_{\left(aq\right)}\\
\rightleftharpoons\mathrm{C_{4.9720}}\mathrm{H}_{7.7882}\mathrm{N}_{1.3240}\mathrm{O}_{1.5231}\mathrm{S}_{0.0441}^{-0.0138}+5.1464\mathrm{O_{2}}_{\left(g\right)}+0.0138\mathrm{H^{+}}\,.\label{react:YHR098C}\end{multline}
The double arrows signify that \emph{a priori }one does not know the
sign of the chemical affinity of either of these reactions.

At 929 residues, YHR098C is over 3 times as long as YLR208W, but in
the formation reactions from the basis species of the residue equivalents
of the two proteins, the coefficients on the basis species are similar.
The difference between the coefficients of the same basis species
in the reactions signifies the response (owing to moderation, i.e.
LeChatelier's principle \cite{Hel87c}) of the metastable equilibrium
assemblage to changes in the corresponding chemical activity or fugacity.
For example, because $\nu_{\mathrm{CO}_{2},\ref{react:YLR208W}}<\nu_{\mathrm{CO_{2}},\ref{react:YHR098C}}$,
$\nu_{\mathrm{NH_{3}},\ref{react:YLR208W}}<\nu_{\mathrm{NH_{3}},\ref{react:YHR098C}}$
and $\nu_{\mathrm{O_{2}},\ref{react:YLR208W}}<\nu_{\mathrm{O_{2}},\ref{react:YHR098C}}$,
increasing $a_{\mathrm{CO_{2}}_{\left(aq\right)}}$, $a_{\mathrm{NH_{3}}_{\left(aq\right)}}$
or $f_{\mathrm{O_{2}}_{\left(g\right)}}$ at constant $T$, $P$ and
chemical activities of the other basis species shifts the metastable
equilibrium in favor of YLR208W at the expense of YHR098C. Here, $\nu_{i}$
denotes the reaction coefficient of the $i$th basis species or protein,
which is negative for reactants and positive for products as written.
Conversely, because $\nu_{\mathrm{H_{2}O},\ref{react:YLR208W}}>\nu_{\mathrm{H_{2}O},\ref{react:YHR098C}}$,
$\nu_{\mathrm{H_{2}S},\ref{react:YLR208W}}>\nu_{\mathrm{H_{2}S},\ref{react:YHR098C}}$
and $\nu_{\mathrm{H^{+}},\ref{react:YLR208W}}>\nu_{\mathrm{H^{+}},\ref{react:YHR098C}}$
increasing $a_{\mathrm{H_{2}O}}$, $a_{\mathrm{H_{2}S}_{\left(aq\right)}}$
or $a_{\mathrm{H^{+}}}$ (decreasing $\mathrm{pH}$) at constant $T$,
$P$ and chemical activities of the other basis species shifts the
metastable equilibrium in favor of YHR098C at the expense of YLR208W.
The magnitude of the effect is proportional to the size of the difference
between the coefficients of the basis species in the reactions, and
it can be quantified for a specific model system using the following
calculations.

To assess the relative abundances of the proteins in metastable equilibrium,
we proceed by calculating the chemical affinities of each of the formation
reactions. The chemical affinity ($\vec{A}$) is calculated by combining
the equilibrium constant ($K$) with the reaction activity product
($Q$) according to \cite{PD54}\begin{equation}
\vec{A}/2.303RT=\log\left(K/Q\right)=\log\left(\frac{-\Delta G_{r}^{\circ}/2.303RT}{\prod a_{i}^{\nu_{i}}}\right)\,,\label{eq:affinity}\end{equation}
where 2.303 is the natural logarithm of 10, $R$ stands for the gas
constant, $T$ is temperature in degrees Kelvin, $\Delta G_{r}^{\circ}$
is the standard molal Gibbs energy of the reaction, and $a_{i}$ and
$\nu_{i}$ represent the chemical activity and reaction coefficient
of the $i$th basis species or species of interest (i.e., residue
equivalent of the protein) in the reaction. Let us calculate $\Delta G_{r}^{\circ}$
(in kcal mol$^{-1}$) of Reaction \ref{react:YLR208W} by writing

\begin{align}
\Delta G_{\ref{react:YLR208W}}^{\circ} & =1\times-36.633+5.1414\times0+0.0366\times0\nonumber \\
 & -5.0000\times-92.250-1.7946\times-56.688\nonumber \\
 & -1.3468\times-6.383-0.0135\times-6.673\nonumber \\
 & =535.036\,.\label{eq:G_YLR208W}\end{align}
In Eqn. (\ref{eq:G_YLR208W}) the values of $\Delta G^{\circ}$ of
$\mathrm{O_{2}}_{\left(g\right)}$ and $\mathrm{H^{+}}$ are both
zero, which are consistent with the standard state conventions for
gases and the hydrogen ion convention used in solution chemistry.
The values of $\Delta G^{\circ}$ of the other basis species are taken
from the literature \cite{HK74a,WEP82,SHS89}. The value of $\log K_{\ref{react:YLR208W}}$
consistent with Eqn. (\ref{eq:G_YLR208W}) is $-392.19$.

We now calculate the activity product of the reaction using\begin{align}
\log Q_{\ref{react:YLR208W}} & =1\times0+5.1414\times-75.3+0.0366\times-7\nonumber \\
 & -5.0000\times-3-1.7946\times0-1.3468\times-4-0.0135\times-7\nonumber \\
 & =-366.92\,.\label{eq:Q_YLR208W}\end{align}
The values of $a_{i}$ used to write Eqn. (\ref{eq:Q_YLR208W}) are
the reference values listed in the Methods for $a_{\mathrm{CO_{2}}_{\left(aq\right)}}$,
$a_{\mathrm{H_{2}O}}$, $a_{\mathrm{NH_{3}}_{\left(aq\right)}}$,
$a_{\mathrm{H_{2}S}_{\left(aq\right)}}$ and $a_{\mathrm{H^{+}}}$.
The value of $f_{\mathrm{O_{2}}_{\left(g\right)}}$ used in Eqn. (\ref{eq:Q_YLR208W})
($\log f_{\mathrm{O_{2}}_{\left(g\right)}}=-75.3$) is also a reference
value that, it will be shown, characterizes a metastable equilibrium
distribution of proteins that is rank-identical to the measured relative
abundances of the proteins. Finally, the value of $a$ of the residue
equivalent of the protein in Eqn. (\ref{eq:Q_YLR208W}) is set to
a reference value of unity ($\log a=0$). If we are only concerned
with the relative abundances of the proteins in metastable equilibrium,
the actual value used here does not matter so long as it is the same
in the analogous calculations for the other proteins.

Combining Eqns. (\ref{eq:affinity})--(\ref{eq:Q_YLR208W}) yields
$\vec{A}_{\ref{react:YLR208W}}/2.303RT=-25.25$ (this is a non-dimensional
number). Following the same procedure for the other four proteins
(YHR098C, YDL195W, YNL049C and YPL085W) results in $\vec{A}/2.303RT$
equal to $-24.86$, $-24.74$, $-24.93$ and $-24.94$, respectively.
Now let us turn to the relative abundances of the proteins in metastable
equilibrium, which we compute using a Boltzmann distribution for the
relative abundances of the residue equivalents:\begin{equation}
\frac{a_{i}}{a_{t}}=\frac{e^{\vec{A}_{i}/RT}}{\sum_{i=1}^{n}e^{\vec{A}_{i}/RT}}\,,\label{eq:Boltzmann}\end{equation}
where $a_{t}$ denotes the total activity of residue equivalents in
the system and $n$ stands for the number of proteins in the system.
Note regarding the left-hand side of Eqn. (\ref{eq:Boltzmann}) that
because we are taking activity coefficients of unity, the ratio $a_{i}/a_{t}$
is equal to the ratio of concentrations, or proportionally numbers,
of residue equivalents in the system. There is not a negative sign
in front of $\vec{A}/RT$ in the exponents Eqn. (\ref{eq:Boltzmann})
because the chemical affinity is the negative of Gibbs energy change
of the reaction. Note in addition that the values of $\vec{A}/2.303RT$
given above must be multiplied by $\ln10=2.303$ before being substituted
in Eqn. (\ref{eq:Boltzmann}). By taking $a_{t}=1$, we can combine
Eqn. (\ref{eq:Boltzmann}) with $\vec{A}/RT$ of each of the formation
reactions to calculate chemical activities of the residue equivalents
of the proteins equal to $0.0905$, $0.2248$, $0.2994$, $0.1944$
and $0.1909$, respectively. The lengths of the proteins are $297$,
$929$, $1273$, $876$ and $2195$, so the corresponding logarithms
of activities of the proteins are e.g. $\log\left(0.0905/297\right)=-3.52$
for YLR208W, and $-3.61$, $-3.63$, $-3.65$ and $-4.06$ for the
remaining proteins, respectively.

If one now iterates calculation of the chemical affinities of the
residue formation reactions using the calculated metastable equilibrium
logarithms of activities of the residue equivalents (instead of the
starting reference value of $\log a=0$), the resulting chemical affinities
for each formation reaction will be all equal and generally non-zero.
This property of metastable equilibrium was used in \cite{Dic08a}
to describe specific application of a method using a system of linear
equations for finding the metastable equilibrium state without explicitly
writing Eqn. (\ref{eq:Boltzmann}).

The results of the calculation described above correspond to the dotted
line at $\log f_{\mathrm{O_{2}}_{\left(g\right)}}=-75.3$ in Fig.
\ref{fig:loga-logfO2}b. At this oxygen fugacity, the ranks of abundance
of the model proteins in metastable equilibrium are identical to the
ranks of experimental abundances. The figure was generated in whole
by carrying out this procedure for different reference values of $\log f_{\mathrm{O_{2}}_{\left(g\right)}}$.
It can be seen in Fig. \ref{fig:loga-logfO2}a that there is a narrow
range on either side of $\log f_{\mathrm{O_{2}}_{\left(g\right)}}=-75.3$
(ca. $\pm0.05$) where the relative abundances of the proteins in
metastable equilibrium occur in the same rank order. Beyond these
limits, changing $f_{\mathrm{O_{2}}_{\left(g\right)}}$ drives the
composition of the metastable equilibrium assemblage to other states
that do not overlap as closely with the experimental rankings. The
experimental abundances of the proteins reported by \cite{GHB+03}
are 21400, 12200, 1840, 1720 and 358, respectively, in relative units.
These abundances were scaled to the same total activity of residues
(unity) used in the calculations to generate the experimental relative
abundances plotted at the dashed line in Fig. \ref{fig:loga-logfO2}b
at $\log f_{\mathrm{O_{2}}_{\left(g\right)}}=-78$. Under these conditions,
the metastable equilibrium abundances of the proteins do not occur
in exactly the same rank order as the experimental ones, but there
is a greater overall correspondence with the experimental relative
abundances.

\subsection*{Relative abundances of proteins within compartments}

\begin{table}
\caption{\label{table:fO2}Oxygen fugacities, deviations and correlation coefficients
in comparisons of intracompartmental protein interactions$^{\mathbf{a}}$.}
\renewcommand{\arraystretch}{0.8}

\begin{centering}
\begin{tabular}{lrrrrrrrrrr}
\hline 
\multicolumn{5}{c}{{\footnotesize Most abundant proteins}} &  & \multicolumn{5}{c}{{\footnotesize Model complexes}}\tabularnewline
\cline{1-5} \cline{7-11} 
{\footnotesize Location} & {\footnotesize $n$} & {\footnotesize $\log f_{\mathrm{O_{2}}_{\left(g\right)}}$} & {\footnotesize RMSD} & $\rho$ &  & {\footnotesize Complex} & {\footnotesize $n$} & {\footnotesize $\log f_{\mathrm{O_{2}}_{\left(g\right)}}$} & {\footnotesize RMSD} & $\rho$\tabularnewline
\hline
{\footnotesize actin} & {\footnotesize 22} & {\footnotesize -75.5} & {\footnotesize 0.61} & {\footnotesize 0.19} &  & {\footnotesize 1} & {\footnotesize 5} & {\footnotesize -80.0} & {\footnotesize 0.35} & {\footnotesize 0.90}\tabularnewline
{\footnotesize ambiguous} & {\footnotesize 50} & {\footnotesize -74.5} & {\footnotesize 0.90} & {\footnotesize 0.16} &  & {\footnotesize 2} & {\footnotesize 7} & {\footnotesize -78.0} & {\footnotesize 0.58} & {\footnotesize 0.57}\tabularnewline
{\footnotesize bud} & {\footnotesize 50} & {\footnotesize -74.5} & {\footnotesize 0.85} & {\footnotesize 0.02} &  & {\footnotesize 3} & {\footnotesize 5} & {\footnotesize -75.0} & {\footnotesize 0.25} & {\footnotesize 0.80}\tabularnewline
{\footnotesize bud.neck} & {\footnotesize 11} & {\footnotesize -75.5} & {\footnotesize 0.73} & {\footnotesize 0.02} &  & {\footnotesize 4} & {\footnotesize 6} & {\footnotesize -75.0} & {\footnotesize 0.80} & {\footnotesize 0.71}\tabularnewline
{\footnotesize cell.periphery} & {\footnotesize 38} & {\footnotesize -74.5} & {\footnotesize 0.63} & {\footnotesize 0.42} &  & {\footnotesize 5} & {\footnotesize 4} & {\footnotesize -74.5} & {\footnotesize 0.29} & {\footnotesize 0.80}\tabularnewline
{\footnotesize cytoplasm} & {\footnotesize 50} & {\footnotesize -77.0} & {\footnotesize 1.20} & {\footnotesize 0.16} &  & {\footnotesize 6} & {\footnotesize 7} & {\footnotesize -80.0} & {\footnotesize 0.20} & {\footnotesize 0.96}\tabularnewline
{\footnotesize early.Golgi} & {\footnotesize 9} & {\footnotesize -74.0} & {\footnotesize 0.72} & {\footnotesize 0.45} &  & {\footnotesize 7} & {\footnotesize 4} & {\footnotesize -75.5} & {\footnotesize 0.21} & {\footnotesize 0.80}\tabularnewline
{\footnotesize endosome} & {\footnotesize 30} & {\footnotesize -75.5} & {\footnotesize 0.98} & {\footnotesize 0.04} &  & {\footnotesize 8} & {\footnotesize 4} & {\footnotesize -73.5} & {\footnotesize 0.60} & {\footnotesize -1.00}\tabularnewline
{\footnotesize ER} & {\footnotesize 49} & {\footnotesize -76.0} & {\footnotesize 0.94} & {\footnotesize 0.09} &  & {\footnotesize 9} & {\footnotesize 3} & {\footnotesize -77.0} & {\footnotesize 0.16} & {\footnotesize -1.00}\tabularnewline
{\footnotesize ER.to.Golgi} & {\footnotesize 5} & {\footnotesize -78.0} & {\footnotesize 0.40} & {\footnotesize 0.40} &  & {\footnotesize 10} & {\footnotesize 4} & {\footnotesize -76.5} & {\footnotesize 0.82} & {\footnotesize -0.80}\tabularnewline
{\footnotesize Golgi} & {\footnotesize 14} & {\footnotesize -75.5} & {\footnotesize 0.80} & {\footnotesize 0.08} &  & {\footnotesize 11} & {\footnotesize 10} & {\footnotesize -74.5} & {\footnotesize 0.76} & {\footnotesize -0.28}\tabularnewline
{\footnotesize late.Golgi} & {\footnotesize 29} & {\footnotesize -74.5} & {\footnotesize 0.61} & {\footnotesize 0.18} &  & {\footnotesize 12} & {\footnotesize 5} & {\footnotesize -76.5} & {\footnotesize 1.74} & {\footnotesize -0.30}\tabularnewline
{\footnotesize lipid.particle} & {\footnotesize 17} & {\footnotesize -78.0} & {\footnotesize 0.92} & {\footnotesize 0.23} &  & {\footnotesize 13} & {\footnotesize 12} & {\footnotesize -74.5} & {\footnotesize 0.97} & {\footnotesize 0.01}\tabularnewline
{\footnotesize microtubule} & {\footnotesize 10} & {\footnotesize -75.0} & {\footnotesize 0.61} & {\footnotesize 0.36} &  & {\footnotesize 14} & {\footnotesize 7} & {\footnotesize -73.0} & {\footnotesize 1.05} & {\footnotesize -0.93}\tabularnewline
{\footnotesize mitochondrion} & {\footnotesize 50} & {\footnotesize -76.0} & {\footnotesize 0.49} & {\footnotesize 0.43} &  & {\footnotesize 15} & {\footnotesize 17} & {\footnotesize -74.0} & {\footnotesize 0.49} & {\footnotesize -0.14}\tabularnewline
{\footnotesize nuclear.periphery} & {\footnotesize 46} & {\footnotesize -76.0} & {\footnotesize 0.62} & {\footnotesize 0.32} &  & {\footnotesize 16} & {\footnotesize 23} & {\footnotesize -76.0} & {\footnotesize 0.43} & {\footnotesize 0.53}\tabularnewline
{\footnotesize nucleolus} & {\footnotesize 50} & {\footnotesize -75.5} & {\footnotesize 0.79} & {\footnotesize -0.18} &  & {\footnotesize 17} & {\footnotesize 6} & {\footnotesize -74.0} & {\footnotesize 0.57} & {\footnotesize 0.66}\tabularnewline
{\footnotesize nucleus} & {\footnotesize 50} & {\footnotesize -75.0} & {\footnotesize 0.80} & {\footnotesize -0.02} &  & {\footnotesize 18} & {\footnotesize 5} & {\footnotesize -79.0} & {\footnotesize 0.25} & {\footnotesize 0.90}\tabularnewline
{\footnotesize peroxisome} & {\footnotesize 18} & {\footnotesize -75.5} & {\footnotesize 0.55} & {\footnotesize 0.56} &  & {\footnotesize 19} & {\footnotesize 8} & {\footnotesize -76.0} & {\footnotesize 0.39} & {\footnotesize 0.91}\tabularnewline
{\footnotesize punctate.composite} & {\footnotesize 49} & {\footnotesize -74.0} & {\footnotesize 0.78} & {\footnotesize 0.19} &  & {\footnotesize 20} & {\footnotesize 15} & {\footnotesize -74.5} & {\footnotesize 0.59} & {\footnotesize 0.66}\tabularnewline
{\footnotesize spindle.pole} & {\footnotesize 30} & {\footnotesize -76.0} & {\footnotesize 1.07} & {\footnotesize -0.13} &  & {\footnotesize 21} & {\footnotesize 5} & {\footnotesize -80.0} & {\footnotesize 1.06} & {\footnotesize 0.60}\tabularnewline
{\footnotesize vacuolar.membrane} & {\footnotesize 45} & {\footnotesize -76.5} & {\footnotesize 1.07} & {\footnotesize 0.36} &  & {\footnotesize 22} & {\footnotesize 15} & {\footnotesize -78.5} & {\footnotesize 1.14} & {\footnotesize 0.14}\tabularnewline
{\footnotesize vacuole} & {\footnotesize 50} & {\footnotesize -74.5} & {\footnotesize 1.49} & {\footnotesize -0.02} &  & {\footnotesize 23} & {\footnotesize 9} & {\footnotesize -78.0} & {\footnotesize 0.93} & {\footnotesize 0.32}\tabularnewline
\hline
\end{tabular}
\par\end{centering}

\textbf{a.} Values of $\log f_{\mathrm{O_{2}}_{\left(g\right)}}$
in each location were regressed by comparing calculated and experimental
logarithms of activities of the most abundant proteins in different
subcellular locations and of selected complexes for each location
(Figure S1). $n$ denotes the number of model proteins used in the
calculations. RMSD values were calculated using Eqn. (\ref{eq:RMSD}),
and $\rho$ denotes the Spearman rank correlation coefficient, calculated
using Eqn. (\ref{eq:spearman}).
\end{table}

\begin{figure}
\begin{centering}
\includegraphics[width=0.95\columnwidth]{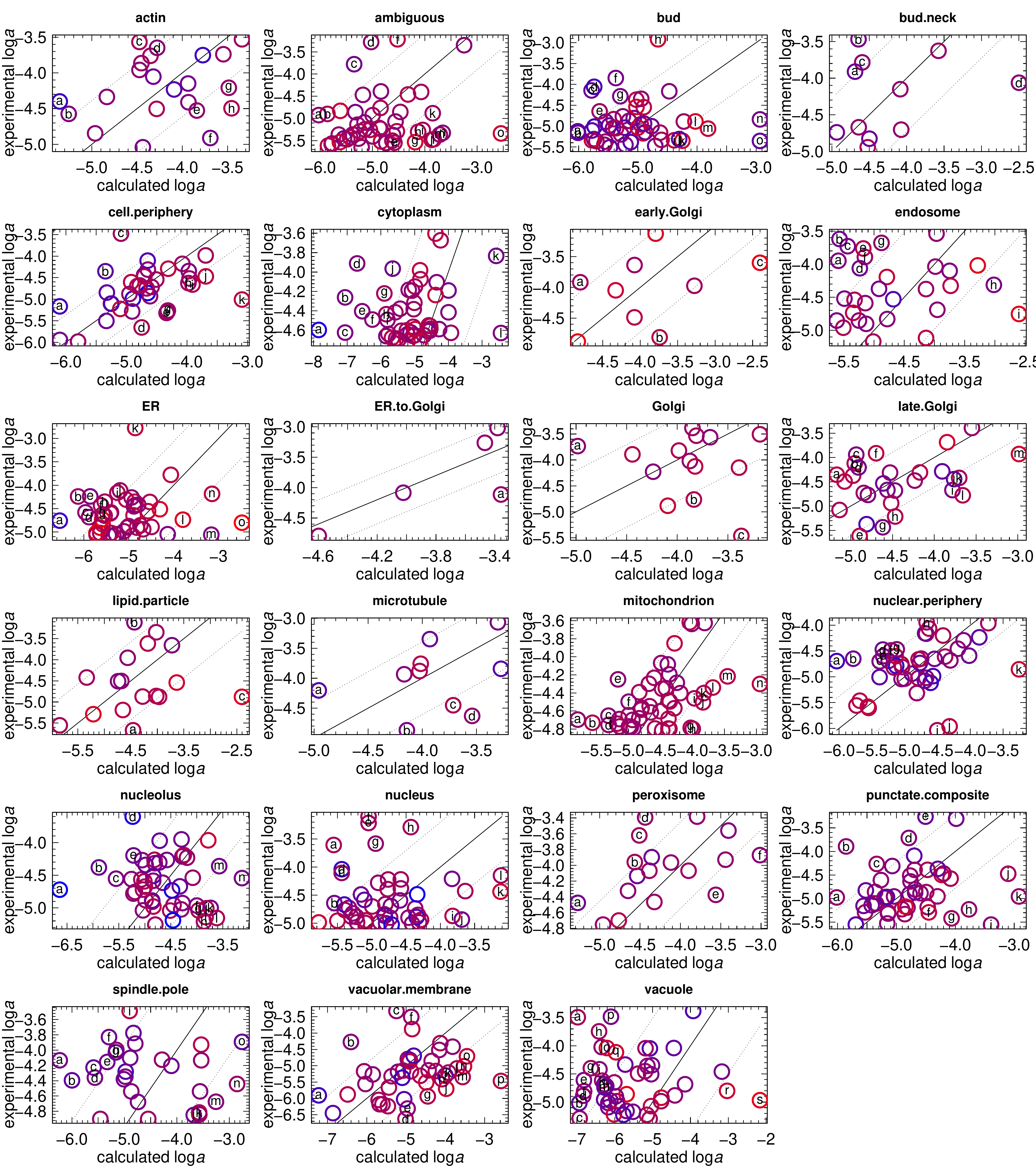}
\par\end{centering}

\caption{\label{fig:proteome_loga}\textbf{Comparison of experimental and calculated
logarithms of activities of proteins in compartments}. Red and blue
colors denote, respectively, low and high average nominal carbon oxidation
states ($\overline{Z}_{\mathrm{C}}$) of the protein. Dotted lines
are positioned at one RMSD above and below one-to-one correspondence,
which is denoted by the solid lines. Outlying points are labeled with
letters that are keyed to the proteins in Table \ref{table:outliers}.
The values of $\log f_{\mathrm{O_{2}}_{\left(g\right)}}$ used in
the calculations are listed in Table \ref{table:fO2}.}

\end{figure}

The procedure outlined above for calculating the relative abundances
of model proteins in ER to Golgi was repeated for each of the other
compartments identified in \cite{HFG+03}. Up to 50 experimentally
most abundant proteins were chosen to model each of the compartments.
The relative abundances of the proteins were calculated at 0.5 $\log$
unit increments from $\log f_{\mathrm{O_{2}}_{\left(g\right)}}=-82$
to $-70.5$. Scatterplots of the experimental vs. calculated relative
abundances for each set of proteins are shown in Figure S1. These
comparisons were visually assessed to regress values of $\log f_{\mathrm{O_{2}}_{\left(g\right)}}$,
listed in Table \ref{table:fO2}, that yield the best fit between
calculated and experimental relative abundances. The resulting calculated
relative abundances are listed together with the experimental ones
in Table S3; the best-fit scatterplots for each set of model proteins
are shown in Fig. \ref{fig:proteome_loga}

The retrieval of optimal values of $\log f_{\mathrm{O}_{2\left(g\right)}}$
was aided by also calculating the root mean square deviation (RMSD)
of logarithms of activities using Eqn. (\ref{eq:RMSD}) and the Spearman
rank correlation coefficient ($\rho$; Eqn. \ref{eq:spearman}) between
experimental and calculated logarithms of activities. The dotted lines
in Fig. \ref{fig:proteome_loga} were drawn at one RMSD on either
side of the one-to-one correspondence, denoted by the solid lines
in this figure. The RMSD values were used to identify outliers that
are identified in Fig. \ref{fig:proteome_loga} by letters and open
symbols and that are listed in Table \ref{table:outliers}. To aid
in distinguishing the points, they were assigned colors on a red-blue
scale that denotes the average nominal oxidation state of carbon of
the protein (Eqn. \ref{eq:ZC}).

There is a considerable degree of scatter apparent in many of the
plots shown in Fig. \ref{fig:proteome_loga}, so a low significance
is attached with the $\log f_{\mathrm{O_{2}}_{\left(g\right)}}$ values
regressed from these comparisons. In specific cases such as late Golgi
and nuclear periphery a lower overall deviation is apparent and there
is a visual indication of a positive correlation between the calculated
and experimental relative abundances. Because they were regressed
from individual noisy data, the values of $\log f_{\mathrm{O_{2}}_{\left(g\right)}}$
listed in Table \ref{table:fO2} are probably not as representative
of subcellular oxidation-reduction conditions as those listed in Table
\ref{table:proteologs}, which have the additional benefit of being
partly based on known subcellular interactions (see above).

\begin{table}
\caption{\label{table:outliers}Outliers in Fig. \ref{fig:proteome_loga}$^{\mathbf{a}}$.}

\renewcommand{\arraystretch}{0.6}

\begin{centering}
\begin{tabular}{ccccllllllll}
\hline 
{\scriptsize ID} & {\scriptsize ORF} & {\scriptsize ID} & {\scriptsize ORF} & {\scriptsize ID} & {\scriptsize ORF} & {\scriptsize ID} & {\scriptsize ORF} & {\scriptsize ID} & {\scriptsize ORF} & {\scriptsize ID} & {\scriptsize ORF}\tabularnewline
\hline
\multicolumn{2}{c}{{\scriptsize actin}} & \multicolumn{2}{c}{{\scriptsize cell.periphery}} & \multicolumn{2}{c}{{\scriptsize endosome}} & \multicolumn{2}{c}{{\scriptsize microtubule}} & \multicolumn{2}{c}{{\scriptsize nucleolus}} & \multicolumn{2}{c}{{\scriptsize spindle.pole}}\tabularnewline
{\scriptsize a} & {\scriptsize YLR206W} & {\scriptsize a} & {\scriptsize YDR034W-B} & {\scriptsize a} & {\scriptsize YBR131W} & {\scriptsize a} & {\scriptsize YBL031W} & {\scriptsize a} & {\scriptsize YKR092C} & {\scriptsize a} & {\scriptsize YLR457C}\tabularnewline
{\scriptsize b} & {\scriptsize YIL095W} & {\scriptsize b} & {\scriptsize YLL010C} & {\scriptsize b} & {\scriptsize YOR132W} & {\scriptsize b} & {\scriptsize YBL063W} & {\scriptsize b} & {\scriptsize YLL011W} & {\scriptsize b} & {\scriptsize YPL255W}\tabularnewline
{\scriptsize c} & {\scriptsize YNR035C} & {\scriptsize c} & {\scriptsize YOR153W} & {\scriptsize c} & {\scriptsize YNR006W} & {\scriptsize c} & {\scriptsize YPL209C} & {\scriptsize c} & {\scriptsize YNL299W} & {\scriptsize c} & {\scriptsize YJR053W}\tabularnewline
{\scriptsize d} & {\scriptsize YMR092C} & {\scriptsize d} & {\scriptsize YBR043C} & {\scriptsize d} & {\scriptsize YMR171C} & {\scriptsize d} & {\scriptsize YMR198W} & {\scriptsize d} & {\scriptsize YGR159C} & {\scriptsize d} & {\scriptsize YDR356W}\tabularnewline
{\scriptsize e} & {\scriptsize YGR080W} & {\scriptsize e} & {\scriptsize YDR038C} & {\scriptsize e} & {\scriptsize YLR240W} &  &  & {\scriptsize e} & {\scriptsize YMR014W} & {\scriptsize e} & {\scriptsize YOR373W}\tabularnewline
{\scriptsize f} & {\scriptsize YCR088W} & {\scriptsize f} & {\scriptsize YDR039C} & {\scriptsize f} & {\scriptsize YLR073C} & \multicolumn{2}{c}{{\scriptsize mitochondrion}} & {\scriptsize f} & {\scriptsize YJR063W} & {\scriptsize f} & {\scriptsize YGL061C}\tabularnewline
{\scriptsize g} & {\scriptsize YIL062C} & {\scriptsize g} & {\scriptsize YDR040C} & {\scriptsize g} & {\scriptsize YGR206W} & {\scriptsize a} & {\scriptsize YIL125W} & {\scriptsize g} & {\scriptsize YGR271C-A} & {\scriptsize g} & {\scriptsize YKL089W}\tabularnewline
{\scriptsize h} & {\scriptsize YOR367W} & {\scriptsize h} & {\scriptsize YHR146W} & {\scriptsize h} & {\scriptsize YKR035W-A} & {\scriptsize b} & {\scriptsize YNL063W} & {\scriptsize h} & {\scriptsize YCR086W} & {\scriptsize h} & {\scriptsize YIL144W}\tabularnewline
 &  & {\scriptsize i} & {\scriptsize YPR156C} & {\scriptsize i} & {\scriptsize YJR044C} & {\scriptsize c} & {\scriptsize YCL009C} & {\scriptsize i} & {\scriptsize YNR004W} & {\scriptsize i} & {\scriptsize YLR381W}\tabularnewline
\multicolumn{2}{c}{{\scriptsize ambiguous}} & {\scriptsize j} & {\scriptsize YLR413W} &  &  & {\scriptsize d} & {\scriptsize YHR051W} & {\scriptsize j} & {\scriptsize YLR367W} & {\scriptsize j} & {\scriptsize YPL233W}\tabularnewline
{\scriptsize a} & {\scriptsize YGL021W} & {\scriptsize k} & {\scriptsize YOR094W} & \multicolumn{2}{c}{{\scriptsize ER}} & {\scriptsize e} & {\scriptsize YOR108W} & {\scriptsize k} & {\scriptsize YDR156W} & {\scriptsize k} & {\scriptsize YOL069W}\tabularnewline
{\scriptsize b} & {\scriptsize YLR454W} &  &  & {\scriptsize a} & {\scriptsize YLR390W-A} & {\scriptsize f} & {\scriptsize YDR232W} & {\scriptsize l} & {\scriptsize YOR310C} & {\scriptsize l} & {\scriptsize YMR117C}\tabularnewline
{\scriptsize c} & {\scriptsize YHR115C} & \multicolumn{2}{c}{{\scriptsize cytoplasm}} & {\scriptsize b} & {\scriptsize YEL002C} & {\scriptsize g} & {\scriptsize YMR083W} & {\scriptsize m} & {\scriptsize YLR221C} & {\scriptsize m} & {\scriptsize YDR016C}\tabularnewline
{\scriptsize d} & {\scriptsize YER070W} & {\scriptsize a} & {\scriptsize YNL255C} & {\scriptsize c} & {\scriptsize YML012W} & {\scriptsize h} & {\scriptsize YKL040C} & {\scriptsize n} & {\scriptsize YNL113W} & {\scriptsize n} & {\scriptsize YDR201W}\tabularnewline
{\scriptsize e} & {\scriptsize YIL065C} & {\scriptsize b} & {\scriptsize YBL027W} & {\scriptsize d} & {\scriptsize YJR131W} & {\scriptsize i} & {\scriptsize YFL018C} &  &  & {\scriptsize o} & {\scriptsize YKR083C}\tabularnewline
{\scriptsize f} & {\scriptsize YJR011C} & {\scriptsize c} & {\scriptsize YBR084C-A} & {\scriptsize e} & {\scriptsize YDR221W} & {\scriptsize j} & {\scriptsize YPL078C} & \multicolumn{2}{c}{{\scriptsize nucleus}} &  & \tabularnewline
{\scriptsize g} & {\scriptsize YPR139C} & {\scriptsize d} & {\scriptsize YER131W} & {\scriptsize f} & {\scriptsize YOR254C} & {\scriptsize k} & {\scriptsize YKL085W} & {\scriptsize a} & {\scriptsize YBR010W} & \multicolumn{2}{c}{{\scriptsize vacuolar.membrane}}\tabularnewline
{\scriptsize h} & {\scriptsize YHR129C} & {\scriptsize e} & {\scriptsize YML026C} & {\scriptsize g} & {\scriptsize YNL258C} & {\scriptsize l} & {\scriptsize YDR298C} & {\scriptsize b} & {\scriptsize YNL251C} & {\scriptsize a} & {\scriptsize YPL180W}\tabularnewline
{\scriptsize i} & {\scriptsize YHR025W} & {\scriptsize f} & {\scriptsize YDL082W} & {\scriptsize h} & {\scriptsize YML013W} & {\scriptsize m} & {\scriptsize YOR142W} & {\scriptsize c} & {\scriptsize YNR053C} & {\scriptsize b} & {\scriptsize YMR160W}\tabularnewline
{\scriptsize j} & {\scriptsize YAR028W} & {\scriptsize g} & {\scriptsize YGL031C} & {\scriptsize i} & {\scriptsize YHR007C} & {\scriptsize n} & {\scriptsize YDL067C} & {\scriptsize d} & {\scriptsize YDR432W} & {\scriptsize c} & {\scriptsize YDL185W}\tabularnewline
{\scriptsize k} & {\scriptsize YBR256C} & {\scriptsize h} & {\scriptsize YDR012W} & {\scriptsize j} & {\scriptsize YHR042W} &  &  & {\scriptsize e} & {\scriptsize YBR009C} & {\scriptsize d} & {\scriptsize YGL006W}\tabularnewline
{\scriptsize l} & {\scriptsize YMR202W} & {\scriptsize i} & {\scriptsize YMR205C} & {\scriptsize k} & {\scriptsize YKL154W} & \multicolumn{2}{c}{{\scriptsize nuclear.periphery}} & {\scriptsize f} & {\scriptsize YNL030W} & {\scriptsize e} & {\scriptsize YLR447C}\tabularnewline
{\scriptsize m} & {\scriptsize YJL034W} & {\scriptsize j} & {\scriptsize YPR035W} & {\scriptsize l} & {\scriptsize YDL128W} & {\scriptsize a} & {\scriptsize YDL088C} & {\scriptsize g} & {\scriptsize YLR153C} & {\scriptsize f} & {\scriptsize YBR127C}\tabularnewline
{\scriptsize n} & {\scriptsize YMR214W} & {\scriptsize k} & {\scriptsize YDR382W} & {\scriptsize m} & {\scriptsize YKL096W-A} & {\scriptsize b} & {\scriptsize YGR202C} & {\scriptsize h} & {\scriptsize YBL002W} & {\scriptsize g} & {\scriptsize YBR207W}\tabularnewline
{\scriptsize o} & {\scriptsize YJR085C} & {\scriptsize l} & {\scriptsize YOL039W} & {\scriptsize n} & {\scriptsize YBR106W} & {\scriptsize c} & {\scriptsize YKR095W} & {\scriptsize i} & {\scriptsize YDR190C} & {\scriptsize h} & {\scriptsize YML121W}\tabularnewline
 &  &  &  & {\scriptsize o} & {\scriptsize YEL027W} & {\scriptsize d} & {\scriptsize YAR002W} & {\scriptsize j} & {\scriptsize YIL021W} & {\scriptsize i} & {\scriptsize YDR486C}\tabularnewline
\multicolumn{2}{c}{{\scriptsize bud}} & \multicolumn{2}{c}{{\scriptsize early.Golgi}} &  &  & {\scriptsize e} & {\scriptsize YPR174C} & {\scriptsize k} & {\scriptsize YDR513W} & {\scriptsize j} & {\scriptsize YML018C}\tabularnewline
{\scriptsize a} & {\scriptsize YDR309C} & {\scriptsize a} & {\scriptsize YGL223C} & \multicolumn{2}{c}{{\scriptsize ER.to.Golgi}} & {\scriptsize f} & {\scriptsize YER105C} & {\scriptsize l} & {\scriptsize YPL028W} & {\scriptsize k} & {\scriptsize YGR163W}\tabularnewline
{\scriptsize b} & {\scriptsize YBL085W} & {\scriptsize b} & {\scriptsize YBL102W} & {\scriptsize a} & {\scriptsize YNL049C} & {\scriptsize g} & {\scriptsize YGL092W} &  &  & {\scriptsize l} & {\scriptsize YBR077C}\tabularnewline
{\scriptsize c} & {\scriptsize YNL278W} & {\scriptsize c} & {\scriptsize YDR100W} &  &  & {\scriptsize h} & {\scriptsize YFR002W} & \multicolumn{2}{c}{{\scriptsize punctate.composite}} & {\scriptsize m} & {\scriptsize YOR332W}\tabularnewline
{\scriptsize d} & {\scriptsize YNR049C} &  &  & \multicolumn{2}{c}{{\scriptsize Golgi}} & {\scriptsize i} & {\scriptsize YGL247W} & {\scriptsize a} & {\scriptsize YAR009C} & {\scriptsize n} & {\scriptsize YOL092W}\tabularnewline
{\scriptsize e} & {\scriptsize YDR166C} & \multicolumn{2}{c}{{\scriptsize lipid.particle}} & {\scriptsize a} & {\scriptsize YDR245W} & {\scriptsize j} & {\scriptsize YLR450W} & {\scriptsize b} & {\scriptsize YGL200C} & {\scriptsize o} & {\scriptsize YOL129W}\tabularnewline
{\scriptsize f} & {\scriptsize YPL032C} & {\scriptsize a} & {\scriptsize YCL005W} & {\scriptsize b} & {\scriptsize YNL041C} & {\scriptsize k} & {\scriptsize YHR133C} & {\scriptsize c} & {\scriptsize YJL186W} & {\scriptsize p} & {\scriptsize YHR039C-A}\tabularnewline
{\scriptsize g} & {\scriptsize YER149C} & {\scriptsize b} & {\scriptsize YML008C} & {\scriptsize c} & {\scriptsize YLR268W} &  &  & {\scriptsize d} & {\scriptsize YNL243W} &  & \tabularnewline
{\scriptsize h} & {\scriptsize YDR033W} & {\scriptsize c} & {\scriptsize YMR148W} &  &  & \multicolumn{2}{c}{{\scriptsize peroxisome}} & {\scriptsize e} & {\scriptsize YGR086C} & \multicolumn{2}{c}{{\scriptsize vacuole}}\tabularnewline
{\scriptsize i} & {\scriptsize YGR191W} &  &  & \multicolumn{2}{c}{{\scriptsize late.Golgi}} & {\scriptsize a} & {\scriptsize YMR204C} & {\scriptsize f} & {\scriptsize YOL044W} & {\scriptsize a} & {\scriptsize YNL326C}\tabularnewline
{\scriptsize j} & {\scriptsize YMR295C} &  &  & {\scriptsize a} & {\scriptsize YDR407C} & {\scriptsize b} & {\scriptsize YKL197C} & {\scriptsize g} & {\scriptsize YER071C} & {\scriptsize b} & {\scriptsize YER123W}\tabularnewline
{\scriptsize k} & {\scriptsize YLR414C} &  &  & {\scriptsize b} & {\scriptsize YJL044C} & {\scriptsize c} & {\scriptsize YLR324W} & {\scriptsize h} & {\scriptsize YNL173C} & {\scriptsize c} & {\scriptsize YBR205W}\tabularnewline
{\scriptsize l} & {\scriptsize YBR054W} &  &  & {\scriptsize c} & {\scriptsize YDR170C} & {\scriptsize d} & {\scriptsize YGL037C} & {\scriptsize i} & {\scriptsize YDR357C} & {\scriptsize d} & {\scriptsize YER001W}\tabularnewline
{\scriptsize m} & {\scriptsize YLL028W} &  &  & {\scriptsize d} & {\scriptsize YBL010C} & {\scriptsize e} & {\scriptsize YDL022W} & {\scriptsize j} & {\scriptsize YBR052C} & {\scriptsize e} & {\scriptsize YDL211C}\tabularnewline
{\scriptsize n} & {\scriptsize YPR124W} &  &  & {\scriptsize e} & {\scriptsize YMR218C} & {\scriptsize f} & {\scriptsize YGL153W} & {\scriptsize k} & {\scriptsize YDR032C} & {\scriptsize f} & {\scriptsize YOR099W}\tabularnewline
{\scriptsize o} & {\scriptsize YOR304C-A} &  &  & {\scriptsize f} & {\scriptsize YGL083W} &  &  &  &  & {\scriptsize g} & {\scriptsize YPL019C}\tabularnewline
 &  &  &  & {\scriptsize g} & {\scriptsize YDR472W} &  &  &  &  & {\scriptsize h} & {\scriptsize YOL088C}\tabularnewline
\multicolumn{2}{c}{{\scriptsize bud.neck}} &  &  & {\scriptsize h} & {\scriptsize YPL259C} &  &  &  &  & {\scriptsize i} & {\scriptsize YBR199W}\tabularnewline
{\scriptsize a} & {\scriptsize YJR092W} &  &  & {\scriptsize i} & {\scriptsize YBR254C} &  &  &  &  & {\scriptsize j} & {\scriptsize YDR483W}\tabularnewline
{\scriptsize b} & {\scriptsize YPL116W} &  &  & {\scriptsize j} & {\scriptsize YLR330W} &  &  &  &  & {\scriptsize k} & {\scriptsize YIL005W}\tabularnewline
{\scriptsize c} & {\scriptsize YHR023W} &  &  & {\scriptsize k} & {\scriptsize YKR068C} &  &  &  &  & {\scriptsize l} & {\scriptsize YLR300W}\tabularnewline
{\scriptsize d} & {\scriptsize YPR188C} &  &  & {\scriptsize l} & {\scriptsize YKL135C} &  &  &  &  & {\scriptsize m} & {\scriptsize YPR159W}\tabularnewline
 &  &  &  & {\scriptsize m} & {\scriptsize YEL048C} &  &  &  &  & {\scriptsize n} & {\scriptsize YPL163C}\tabularnewline
 &  &  &  &  &  &  &  &  &  & {\scriptsize o} & {\scriptsize YJR161C}\tabularnewline
 &  &  &  &  &  &  &  &  &  & {\scriptsize p} & {\scriptsize YHR215W}\tabularnewline
 &  &  &  &  &  &  &  &  &  & {\scriptsize q} & {\scriptsize YNL336W}\tabularnewline
 &  &  &  &  &  &  &  &  &  & {\scriptsize r} & {\scriptsize YBR187W}\tabularnewline
 &  &  &  &  &  &  &  &  &  & {\scriptsize s} & {\scriptsize YGR105W}\tabularnewline
\hline
\end{tabular}
\par\end{centering}

\textbf{a.} Proteins are listed whose calculated logarithm of activity
differs from experimental values by more than the root mean square
deviation shown in Table \ref{table:fO2}.
\end{table}

The comparisons depicted in Fig. \ref{fig:proteome_loga} and in Figure
S1 are important because they reveal that the range of protein abundance
observed in cells is accessible in a metastable equilibrium assemblage
at some values of $\log f_{\mathrm{O_{2}}_{\left(g\right)}}$. For
example, the range of experimental abundances of the model proteins
in actin covers about $1.6$ orders of magnitude, while the calculated
abundances vary over about $2.2$ orders of magnitude. Extreme values
of $\log f_{\mathrm{O_{2}}_{\left(g\right)}}$ tend to weaken this
correspondence (Figure S1). The lowest degree of correspondence occurs
for the cytoplasmic proteins, where $\sim6$ orders of magnitude separate
the predicted relative abundances of the top $50$ most abundant proteins,
which in the experiments have a dynamic range spanning about $1.2$
orders of magnitude. The great degree of scatter apparent in many
of the comparisons in Fig. \ref{fig:proteome_loga}a is troublesome.
The scatter could be partly a consequence of including in the comparisons
model proteins that do not actually interact with each other, despite
their high relative abundances. To address this concern, a more selective
approach was adopted below that takes account of fewer numbers of
proteins that interact through the formation of complexes.

\subsection*{Relative abundances of proteins in complexes}

The correspondence between the calculated and experimental relative
abundances of the five model proteins in ER to Golgi raises the question
of what characteristics of the proteins might be responsible for this
result. Searching the functional annotations of these proteins reveals
that they are part of the COPII coat complex \cite{SGD}. The inclusion
of the COPII complex above was largely unintentional, as the procedure
there was to look at the most abundant proteins in given compartments.
Nevertheless, the results for that model system suggested that focusing
on specific complexes in other compartments could yield interesting
results. Because the interactions of proteins to form complexes is
essential in cellular structure and regulating functions of enzymes
\cite{DMT05}, factors that affect the relative abundances of the
complexing proteins may be fundamental to the control of metabolic
processes.

\begin{table}
\caption{\label{table:complexes}Model proteins in complexes$^{\mathbf{a}}$.}

\renewcommand{\arraystretch}{0.5}

\begin{centering}
\begin{tabular}{lrrllrrllrrlllrl}
\hline 
{\scriptsize Name} &  & {\scriptsize ORF} &  & {\scriptsize Name} &  & {\scriptsize ORF} &  & {\scriptsize Name} &  & {\scriptsize ORF} &  & {\scriptsize Name} &  & {\scriptsize ORF} & \tabularnewline
\hline
\multicolumn{4}{l}{{\scriptsize 1. actin: Arp2/3 complex (423) }} & \multicolumn{4}{l}{{\scriptsize 9. ER: signal recognition}} & \multicolumn{4}{l}{{\scriptsize 14. microtubule: DASH }} & \multicolumn{4}{l}{{\scriptsize 20. punctate.composite: proteins}}\tabularnewline
\multicolumn{4}{l}{{\scriptsize (\cite{WIM97};}} & \multicolumn{4}{l}{{\scriptsize complex (52)}} & \multicolumn{4}{l}{{\scriptsize complex \cite{MWS+05}}} & \multicolumn{4}{l}{{\scriptsize localized here and early.Golgi}}\tabularnewline
\multicolumn{4}{l}{{\scriptsize \cite{MP99})}} & {\scriptsize Sec65} &  & {\scriptsize YML105C} & {\scriptsize NA} & {\scriptsize Dam1} &  & {\scriptsize YGR113W} & {\scriptsize X} & {\scriptsize Arl1} &  & {\scriptsize YBR164C} & {\scriptsize d}\tabularnewline
{\scriptsize Arc15} &  & {\scriptsize YIL062C} & {\scriptsize b} & {\scriptsize Srp14} &  & {\scriptsize YDL092W} &  & {\scriptsize Duo1} & {\scriptsize {*}} & {\scriptsize YGL061C} & {\scriptsize a} & {\scriptsize Apm3} &  & {\scriptsize YBR288C} & \tabularnewline
{\scriptsize Arc18} &  & {\scriptsize YLR370C} &  & {\scriptsize Srp54} &  & {\scriptsize YPR088C} & {\scriptsize X} & {\scriptsize Dad1} & {\scriptsize {*}} & {\scriptsize YDR016C} &  & {\scriptsize Bug1} &  & {\scriptsize YDL099W} & \tabularnewline
{\scriptsize Arc19} &  & {\scriptsize YKL013C} &  & {\scriptsize Spp68} & {\scriptsize {*}} & {\scriptsize YPL243W} & {\scriptsize a} & {\scriptsize Dad2} & {\scriptsize {*}} & {\scriptsize YKR083C} & {\scriptsize b} & {\scriptsize Arf1} &  & {\scriptsize YDL192W} & \tabularnewline
{\scriptsize Arc35} &  & {\scriptsize YNR035C} & {\scriptsize a} & {\scriptsize Srp72} &  & {\scriptsize YPL210C} & {\scriptsize b} & {\scriptsize Spc19} & {\scriptsize {*}} & {\scriptsize YDR201W} &  & {\scriptsize Luv1} &  & {\scriptsize YDR027C} & {\scriptsize a}\tabularnewline
\cline{5-8} 
{\scriptsize Arc40} &  & {\scriptsize YBR234C} & {\scriptsize NA} & \multicolumn{4}{l}{{\scriptsize 10. ER.to.Golgi: coatomer}} & {\scriptsize Spc34} & {\scriptsize {*}} & {\scriptsize YKR037C} & {\scriptsize NA} & {\scriptsize Tvp23} &  & {\scriptsize YDR084C} & \tabularnewline
{\scriptsize Arp2} & {\scriptsize {*}} & {\scriptsize YDL029W} &  & \multicolumn{4}{l}{{\scriptsize COPII complex (340)}} & {\scriptsize Ask1} & {\scriptsize {*}} & {\scriptsize YKL052C} &  & {\scriptsize Dop1} &  & {\scriptsize YDR141C} & {\scriptsize a}\tabularnewline
{\scriptsize Arp3} &  & {\scriptsize YJR065C} & {\scriptsize X} & {\scriptsize Sec13} &  & {\scriptsize YLR208W} & {\scriptsize a} & {\scriptsize Dad3} & {\scriptsize {*}} & {\scriptsize YBR233W-A} &  & {\scriptsize Kei1} &  & {\scriptsize YDR367W} & \tabularnewline
\cline{1-4} 
\multicolumn{4}{l}{{\scriptsize 2. ambiguous: cyclin-dependent}} & {\scriptsize Sec16} &  & {\scriptsize YPL085W} &  & {\scriptsize Dad4} & {\scriptsize {*}} & {\scriptsize YDR320C-A} &  & {\scriptsize Vrg4} &  & {\scriptsize YGL225W} & \tabularnewline
\multicolumn{4}{l}{{\scriptsize protein kinase complex (343)}} & {\scriptsize Sec23} &  & {\scriptsize YPR181C} & {\scriptsize X} & {\scriptsize Hsk3} &  & {\scriptsize YKL138C-A} & {\scriptsize X} & {\scriptsize Apl6} &  & {\scriptsize YGR261C} & \tabularnewline
\cline{9-12} 
{\scriptsize Cdc28} & {\scriptsize {*}} & {\scriptsize YBR160W} & {\scriptsize b} & {\scriptsize Sfb2} &  & {\scriptsize YNL049C} &  & \multicolumn{4}{l}{{\scriptsize 16. nuclear.periphery: nuclear}} & {\scriptsize Aps3} &  & {\scriptsize YJL024C} & {\scriptsize c}\tabularnewline
{\scriptsize Cks1} & {\scriptsize {*}} & {\scriptsize YBR135W} & {\scriptsize a} & {\scriptsize Sec24} &  & {\scriptsize YIL109C} & {\scriptsize NA} & \multicolumn{4}{l}{{\scriptsize pore complex \cite{RAS+00}}} & {\scriptsize Vps53} &  & {\scriptsize YJL029C} & {\scriptsize NA}\tabularnewline
{\scriptsize Cln2} & {\scriptsize {*}} & {\scriptsize YPL256C} &  & {\scriptsize Grh1} & {\scriptsize {*}} & {\scriptsize YDR517W} &  & {\scriptsize Nup60} &  & {\scriptsize YAR002W} &  & {\scriptsize Tvp38} &  & {\scriptsize YKR088C} & \tabularnewline
\cline{5-8} 
{\scriptsize Cys4} & {\scriptsize {*}} & {\scriptsize YGR155W} &  & \multicolumn{4}{l}{{\scriptsize 11. Golgi: Golgi transport}} & {\scriptsize Nup170} &  & {\scriptsize YBL079W} &  & {\scriptsize Ssp120} &  & {\scriptsize YLR250W} & \tabularnewline
{\scriptsize Sic1} & {\scriptsize {*}} & {\scriptsize YLR079W} &  & \multicolumn{4}{l}{{\scriptsize complex (293)}} & {\scriptsize Asm4} &  & {\scriptsize YDL088C} & {\scriptsize a} & {\scriptsize NA} &  & {\scriptsize YMR010W} & \tabularnewline
{\scriptsize Clb3} & {\scriptsize {*}} & {\scriptsize YDL155W} &  & {\scriptsize Cog1} & {\scriptsize {*}} & {\scriptsize YGL223C} &  & {\scriptsize Nup84} &  & {\scriptsize YDL116W} &  & {\scriptsize NA} &  & {\scriptsize YMR253C} & {\scriptsize NA}\tabularnewline
{\scriptsize Cln1} & {\scriptsize {*}} & {\scriptsize YMR199W} &  & {\scriptsize Cog2} &  & {\scriptsize YGR120C} & {\scriptsize b} & {\scriptsize Gle1} &  & {\scriptsize YDL207W} &  & {\scriptsize Kex2} &  & {\scriptsize YNL238W} & {\scriptsize NA}\tabularnewline
\cline{1-4} 
\multicolumn{4}{l}{{\scriptsize 3. bud: actin-associated motor}} & {\scriptsize Cog3} &  & {\scriptsize YER157W} &  & {\scriptsize Nup42} &  & {\scriptsize YDR192C} & {\scriptsize X} & {\scriptsize Mon2} &  & {\scriptsize YNL297C} & \tabularnewline
\cline{13-16} 
\multicolumn{4}{l}{{\scriptsize protein complex 2 (49) }} & {\scriptsize Cog4} & {\scriptsize {*}} & {\scriptsize YPR105C} &  & {\scriptsize Nup157} &  & {\scriptsize YER105C} & {\scriptsize c} & \multicolumn{4}{l}{{\scriptsize 21. spindle.pole: spindle-pole}}\tabularnewline
\multicolumn{4}{l}{{\scriptsize \cite{SJD+06}}} & {\scriptsize Cog5} &  & {\scriptsize YNL051W} & {\scriptsize c} & {\scriptsize Gle2} &  & {\scriptsize YER107C} &  & \multicolumn{4}{l}{{\scriptsize body complex (219) \cite{VKH+02}}}\tabularnewline
{\scriptsize Myo2} &  & {\scriptsize YAL029C} &  & {\scriptsize Cog6} &  & {\scriptsize YNL041C} &  & {\scriptsize Nic96} &  & {\scriptsize YFR002W} & {\scriptsize g} & {\scriptsize Pfk1} & {\scriptsize {*}} & {\scriptsize YGR240C} & {\scriptsize a}\tabularnewline
{\scriptsize She4} & {\scriptsize {*}} & {\scriptsize YKL130C} & {\scriptsize b} & {\scriptsize Cog7} &  & {\scriptsize YGL005C} &  & {\scriptsize Nup145} &  & {\scriptsize YGL092W} &  & {\scriptsize Spc72} &  & {\scriptsize YAL047C} & \tabularnewline
{\scriptsize Mlc1} &  & {\scriptsize YBR130C} &  & {\scriptsize Cog8} & {\scriptsize {*}} & {\scriptsize YML071C} &  & {\scriptsize Seh1} &  & {\scriptsize YGL100W} & {\scriptsize X} & {\scriptsize Spc97} &  & {\scriptsize YHR172W} & \tabularnewline
{\scriptsize Myo1} &  & {\scriptsize YGL106W} & {\scriptsize X} & {\scriptsize Iml1} & {\scriptsize {*}} & {\scriptsize YJR138W} &  & {\scriptsize Nup49} &  & {\scriptsize YGL172W} & {\scriptsize j} & {\scriptsize Spc98} &  & {\scriptsize YNL126W} & \tabularnewline
{\scriptsize Cmd1} & {\scriptsize {*}} & {\scriptsize YKL007W} &  & {\scriptsize Nrp1} & {\scriptsize {*}} & {\scriptsize YDL167C} & {\scriptsize a} & {\scriptsize Nup57} &  & {\scriptsize YGR119C} &  & {\scriptsize Tub4} &  & {\scriptsize YLR212C} & {\scriptsize b}\tabularnewline
\cline{5-8} \cline{13-16} 
{\scriptsize Myo5} & {\scriptsize {*}} & {\scriptsize YIL034C} & {\scriptsize a} & \multicolumn{4}{l}{{\scriptsize 12. late.Golgi: retrograde }} & {\scriptsize Nup159} &  & {\scriptsize YIL115C} &  & \multicolumn{4}{l}{{\scriptsize 22. vacuolar.membrane: VO}}\tabularnewline
\cline{1-4} 
\multicolumn{4}{l}{{\scriptsize 4. bud.neck: septin complex (333) }} & \multicolumn{4}{l}{{\scriptsize protein complex (114) }} & {\scriptsize Nup192} &  & {\scriptsize YJL039C} &  & \multicolumn{4}{l}{{\scriptsize vacuolar ATPase complex (14)}}\tabularnewline
\multicolumn{4}{l}{{\scriptsize \cite{FWL+98}}} & \multicolumn{4}{l}{{\scriptsize \cite{CCS03}}} & {\scriptsize Nsp1} &  & {\scriptsize YJL041W} & {\scriptsize i} & {\scriptsize Emi2} & {\scriptsize {*}} & {\scriptsize YDR516C} & \tabularnewline
{\scriptsize Bud4} &  & {\scriptsize YJR092W} & {\scriptsize a} & {\scriptsize Kar2} & {\scriptsize {*}} & {\scriptsize YJL034W} & {\scriptsize c} & {\scriptsize Nup82} &  & {\scriptsize YJL061W} & {\scriptsize f} & {\scriptsize Vma6} &  & {\scriptsize YLR447C} & \tabularnewline
{\scriptsize Cdc10} &  & {\scriptsize YCR002C} & {\scriptsize c} & {\scriptsize Vps52} & {\scriptsize {*}} & {\scriptsize YDR484W} &  & {\scriptsize Nup85} &  & {\scriptsize YJR042W} &  & {\scriptsize Vph2} & {\scriptsize {*}} & {\scriptsize YKL119C} & {\scriptsize a}\tabularnewline
{\scriptsize Cdc11} &  & {\scriptsize YJR076C} &  & {\scriptsize Vps53} & {\scriptsize {*}} & {\scriptsize YJL029C} & {\scriptsize NA} & {\scriptsize Nup120} &  & {\scriptsize YKL057C} & {\scriptsize X} & {\scriptsize Bni1} & {\scriptsize {*}} & {\scriptsize YNL271C} & {\scriptsize b}\tabularnewline
{\scriptsize Cdc12} &  & {\scriptsize YHR107C} &  & {\scriptsize Vps54} & {\scriptsize {*}} & {\scriptsize YDR027C} & {\scriptsize a} & {\scriptsize Nup100} &  & {\scriptsize YKL068W} & {\scriptsize h} & {\scriptsize Drs2} & {\scriptsize {*}} & {\scriptsize YAL026C} & \tabularnewline
{\scriptsize Cdc3} &  & {\scriptsize YLR314C} & {\scriptsize X} & {\scriptsize Vps51} & {\scriptsize {*}} & {\scriptsize YKR020W} & {\scriptsize b} & {\scriptsize Nup133} &  & {\scriptsize YKR082W} &  & {\scriptsize Gaa1} & {\scriptsize {*}} & {\scriptsize YLR088W} & {\scriptsize NA}\tabularnewline
{\scriptsize Shs1} &  & {\scriptsize YDL225W} & {\scriptsize b} & {\scriptsize Scj1} & {\scriptsize {*}} & {\scriptsize YMR214W} &  & {\scriptsize Pom34} &  & {\scriptsize YLR018C} & {\scriptsize NA} & {\scriptsize Lys9} & {\scriptsize {*}} & {\scriptsize YNR050C} & \tabularnewline
\cline{5-8} 
{\scriptsize Mdh1} & {\scriptsize {*}} & {\scriptsize YKL085W} &  & \multicolumn{4}{l}{{\scriptsize 13. lipid.particle: sterol }} & {\scriptsize Ndc1} &  & {\scriptsize YML031W} &  & {\scriptsize Nop6} & {\scriptsize {*}} & {\scriptsize YDL213C} & {\scriptsize c}\tabularnewline
\cline{1-4} 
\multicolumn{4}{l}{{\scriptsize 5. cell.periphery: exocyst }} & \multicolumn{4}{l}{{\scriptsize biosynthesis enzymes}} & {\scriptsize Nup188} &  & {\scriptsize YML103C} & {\scriptsize e} & {\scriptsize Pdc1} & {\scriptsize {*}} & {\scriptsize YLR044C} & \tabularnewline
\multicolumn{4}{l}{{\scriptsize complex (120)}} & \multicolumn{4}{l}{{\scriptsize \cite{MB05}}} & {\scriptsize Nup116} & {\scriptsize {*}} & {\scriptsize YMR047C} & {\scriptsize NA} & {\scriptsize Pgi1} & {\scriptsize {*}} & {\scriptsize YBR196C} & \tabularnewline
{\scriptsize Exo84} &  & {\scriptsize YBR102C} & {\scriptsize NA} & {\scriptsize Erg9} & {\scriptsize {*}} & {\scriptsize YHR190W} &  & {\scriptsize Pom152} &  & {\scriptsize YMR129W} & {\scriptsize b} & {\scriptsize Vac8} &  & {\scriptsize YEL013W} & \tabularnewline
{\scriptsize Sec10} &  & {\scriptsize YLR166C} &  & {\scriptsize Erg1} & {\scriptsize {*}} & {\scriptsize YGR175C} &  & {\scriptsize Nup53} &  & {\scriptsize YMR153W} &  & {\scriptsize Vma10} &  & {\scriptsize YHR039C-A} & {\scriptsize d}\tabularnewline
{\scriptsize Sec3} &  & {\scriptsize YER008C} & {\scriptsize b} & {\scriptsize Erg7} &  & {\scriptsize YHR072W} & {\scriptsize c} & {\scriptsize Nup1} &  & {\scriptsize YOR098C} & {\scriptsize d} & {\scriptsize Vma2} &  & {\scriptsize YBR127C} & \tabularnewline
{\scriptsize Sec5} &  & {\scriptsize YDR166C} & {\scriptsize a} & {\scriptsize Erg11} & {\scriptsize {*}} & {\scriptsize YHR007C} &  & {\scriptsize Cdc31} &  & {\scriptsize YOR257W} & {\scriptsize X} & {\scriptsize Vma7} &  & {\scriptsize YGR020C} & \tabularnewline
\cline{9-12} 
{\scriptsize Sec6} &  & {\scriptsize YIL068C} &  & {\scriptsize Erg24} & {\scriptsize {*}} & {\scriptsize YNL280C} &  & \multicolumn{4}{l}{{\scriptsize 17. nucleolus: small subunit}} & {\scriptsize Vph1} &  & {\scriptsize YOR270C} & \tabularnewline
{\scriptsize Sec8} &  & {\scriptsize YPR055W} & {\scriptsize NA} & {\scriptsize Erg25} & {\scriptsize {*}} & {\scriptsize YGR060W} &  & \multicolumn{4}{l}{{\scriptsize processome (70)}} & {\scriptsize Vtc4} &  & {\scriptsize YJL012C} & {\scriptsize X}\tabularnewline
\cline{1-4} 
\multicolumn{4}{l}{{\scriptsize 6. cytoplasm: translation}} & {\scriptsize Erg26} & {\scriptsize {*}} & {\scriptsize YGL001C} &  & \multicolumn{4}{l}{{\scriptsize \cite{BGM+04}}} & {\scriptsize Yor1} & {\scriptsize {*}} & {\scriptsize YGR281W} & \tabularnewline
\multicolumn{4}{l}{{\scriptsize initiation factor eIF3 (45)}} & {\scriptsize Erg27} &  & {\scriptsize YLR100W} & {\scriptsize NA} & {\scriptsize Utp8} &  & {\scriptsize YGR128C} &  & {\scriptsize Yra1} & {\scriptsize {*}} & {\scriptsize YDR381W} & {\scriptsize NA}\tabularnewline
\cline{13-16} 
{\scriptsize Fun12} &  & {\scriptsize YAL035W} &  & {\scriptsize Erg6} &  & {\scriptsize YML008C} & {\scriptsize d} & {\scriptsize Nan1} &  & {\scriptsize YPL126W} & {\scriptsize b} & \multicolumn{4}{l}{{\scriptsize 23. vacuole: vacuolar proteases}}\tabularnewline
{\scriptsize Hcr1} &  & {\scriptsize YLR192C} & {\scriptsize c} & {\scriptsize Erg2} & {\scriptsize {*}} & {\scriptsize YMR202W} &  & {\scriptsize Utp10} &  & {\scriptsize YJL109C} & {\scriptsize a} & \multicolumn{4}{l}{{\scriptsize and other canonical proteins}}\tabularnewline
{\scriptsize Nip1} &  & {\scriptsize YMR309C} & {\scriptsize a} & {\scriptsize Erg3} & {\scriptsize {*}} & {\scriptsize YLR056W} & {\scriptsize a} & {\scriptsize Utp15} &  & {\scriptsize YMR093W} &  & \multicolumn{4}{l}{{\scriptsize \cite{SCC+07}}}\tabularnewline
{\scriptsize Prt1} &  & {\scriptsize YOR361C} &  & {\scriptsize Erg5} & {\scriptsize {*}} & {\scriptsize YMR015C} &  & {\scriptsize Utp4} &  & {\scriptsize YDR324C} &  & {\scriptsize Ape1} & {\scriptsize {*}} & {\scriptsize YKL103C} & {\scriptsize b}\tabularnewline
{\scriptsize Rli1} &  & {\scriptsize YDR091C} &  & {\scriptsize Erg4} & {\scriptsize {*}} & {\scriptsize YGL012W} & {\scriptsize b} & {\scriptsize Utp9} &  & {\scriptsize YHR196W} &  & {\scriptsize Ape3} & {\scriptsize {*}} & {\scriptsize YBR286W} & \tabularnewline
\cline{5-12} 
{\scriptsize Rpg1} &  & {\scriptsize YBR079C} &  & \multicolumn{4}{l}{{\scriptsize 15. mitochondrion: mitochondrial}} & \multicolumn{4}{l}{{\scriptsize 18. nucleus: RNA }} & {\scriptsize Lap3} & {\scriptsize {*}} & {\scriptsize YNL239W} & \tabularnewline
{\scriptsize Tif34} &  & {\scriptsize YMR146C} & {\scriptsize X} & \multicolumn{4}{l}{{\scriptsize ribosome small subunit (9)}} & \multicolumn{4}{l}{{\scriptsize polymerase I (30)}} & {\scriptsize Pep4} &  & {\scriptsize YPL154C} & {\scriptsize NA}\tabularnewline
{\scriptsize Tif35} &  & {\scriptsize YDR429C} & {\scriptsize NA} & {\scriptsize Ehd3} &  & {\scriptsize YDR036C} & {\scriptsize f} & {\scriptsize Rpa49} & {\scriptsize {*}} & {\scriptsize YNL248C} & {\scriptsize NA} & {\scriptsize Prb1} & {\scriptsize {*}} & {\scriptsize YEL060C} & \tabularnewline
{\scriptsize Tif5} &  & {\scriptsize YPR041W} & {\scriptsize b} & {\scriptsize Mrp13} &  & {\scriptsize YGR084C} & {\scriptsize a} & {\scriptsize Rpa12} & {\scriptsize {*}} & {\scriptsize YJR063W} &  & {\scriptsize Prb1} &  & {\scriptsize YMR297W} & \tabularnewline
\cline{1-4} 
\multicolumn{4}{l}{{\scriptsize 7. early.Golgi: SNARE complex}} & {\scriptsize Mrp17} &  & {\scriptsize YKL003C} & {\scriptsize NA} & {\scriptsize Rpa190} & {\scriptsize {*}} & {\scriptsize YOR341W} &  & {\scriptsize Ams1} & {\scriptsize {*}} & {\scriptsize YGL156W} & {\scriptsize a}\tabularnewline
\multicolumn{4}{l}{{\scriptsize (113) \cite{BL04}}} & {\scriptsize Mrp21} &  & {\scriptsize YBL090W} & {\scriptsize h} & {\scriptsize RPApa3} & {\scriptsize {*}} & {\scriptsize YOR340C} &  & {\scriptsize Ath1} &  & {\scriptsize YPR026W} & {\scriptsize X}\tabularnewline
{\scriptsize Dsl1} & {\scriptsize {*}} & {\scriptsize YNL258C} & {\scriptsize a} & {\scriptsize Mrp4} &  & {\scriptsize YHL004W} &  & {\scriptsize Rpc40} &  & {\scriptsize YPR110C} & {\scriptsize a} & {\scriptsize Pho8} &  & {\scriptsize YDR481C} & \tabularnewline
{\scriptsize Sec39} & {\scriptsize {*}} & {\scriptsize YLR440C} &  & {\scriptsize Mrp51} &  & {\scriptsize YPL118W} &  & {\scriptsize Rpa135} & {\scriptsize {*}} & {\scriptsize YPR010C} &  & {\scriptsize Vtc4} &  & {\scriptsize YJL012C} & {\scriptsize X}\tabularnewline
{\scriptsize Tip20} & {\scriptsize {*}} & {\scriptsize YGL145W} &  & {\scriptsize Mrps16} &  & {\scriptsize YPL013C} & {\scriptsize NA} & {\scriptsize Rpb5} &  & {\scriptsize YBR154C} & {\scriptsize X} & {\scriptsize Ypt7} & {\scriptsize {*}} & {\scriptsize YML001W} & {\scriptsize c}\tabularnewline
\cline{9-12} 
{\scriptsize Ufe1} & {\scriptsize {*}} & {\scriptsize YOR075W} & {\scriptsize NA} & {\scriptsize Mrps17} &  & {\scriptsize YMR188C} & {\scriptsize c} & \multicolumn{4}{l}{{\scriptsize 19. peroxisome: integral to}} & {\scriptsize Npc2} &  & {\scriptsize YDL046W} & {\scriptsize d}\tabularnewline
{\scriptsize Use1} & {\scriptsize {*}} & {\scriptsize YGL098W} &  & {\scriptsize Mrps18} &  & {\scriptsize YNL306W} &  & \multicolumn{4}{l}{{\scriptsize peroxisomal membrane}} & {\scriptsize NA} &  & {\scriptsize YHR202W} & {\scriptsize NA}\tabularnewline
\cline{13-16} 
{\scriptsize Pep12} &  & {\scriptsize YOR036W} & {\scriptsize X} & {\scriptsize Mrps28} &  & {\scriptsize YDR337W} & {\scriptsize e} & \multicolumn{4}{l}{{\scriptsize (GO:0005779)}} &  &  &  & \tabularnewline
{\scriptsize Ykt6} &  & {\scriptsize YKL196C} & {\scriptsize X} & {\scriptsize Mrps5} &  & {\scriptsize YBR251W} & {\scriptsize d} & {\scriptsize Ant1} &  & {\scriptsize YPR128C} &  &  &  &  & \tabularnewline
\cline{1-4} 
\multicolumn{4}{l}{{\scriptsize 8. endosome: ESCRT I \& II }} & {\scriptsize Mrps8} &  & {\scriptsize YMR158W} & {\scriptsize X} & {\scriptsize Inp2} & {\scriptsize {*}} & {\scriptsize YMR163C} &  &  &  &  & \tabularnewline
\multicolumn{4}{l}{{\scriptsize complexes (\cite{KST+07};}} & {\scriptsize Mrps9} &  & {\scriptsize YBR146W} & {\scriptsize X} & {\scriptsize Pex12} &  & {\scriptsize YMR026C} &  &  &  &  & \tabularnewline
\multicolumn{4}{l}{{\scriptsize \cite{HSR+04})}} & {\scriptsize Pet123} &  & {\scriptsize YOR158W} & {\scriptsize g} & {\scriptsize Pex15} & {\scriptsize {*}} & {\scriptsize YOL044W} &  &  &  &  & \tabularnewline
{\scriptsize Vps23} &  & {\scriptsize YCL008C} & {\scriptsize X} & {\scriptsize Rsm10} &  & {\scriptsize YDR041W} & {\scriptsize b} & {\scriptsize Pex22} & {\scriptsize {*}} & {\scriptsize YAL055W} & {\scriptsize b} &  &  &  & \tabularnewline
{\scriptsize Vps28} & {\scriptsize {*}} & {\scriptsize YPL065W} &  & {\scriptsize Rsm19} &  & {\scriptsize YNR037C} & {\scriptsize X} & {\scriptsize Pex3} &  & {\scriptsize YDR329C} &  &  &  &  & \tabularnewline
{\scriptsize Vps37} &  & {\scriptsize YLR119W} & {\scriptsize X} & {\scriptsize Rsm22} &  & {\scriptsize YKL155C} &  & {\scriptsize Pex30} &  & {\scriptsize YLR324W} & {\scriptsize c} &  &  &  & \tabularnewline
{\scriptsize Mvb12} &  & {\scriptsize YGR206W} & {\scriptsize a} & {\scriptsize Rsm23} &  & {\scriptsize YGL129C} &  & {\scriptsize Pex31} & {\scriptsize {*}} & {\scriptsize YGR004W} & {\scriptsize a} &  &  &  & \tabularnewline
{\scriptsize Vps22} & {\scriptsize {*}} & {\scriptsize YPL002C} & {\scriptsize b} & {\scriptsize Rsm27} &  & {\scriptsize YGR215W} & {\scriptsize NA} & {\scriptsize Pex32} & {\scriptsize {*}} & {\scriptsize YBR168W} & {\scriptsize NA} &  &  &  & \tabularnewline
{\scriptsize Vps36} & {\scriptsize {*}} & {\scriptsize YLR417W} &  & {\scriptsize Rsm7} &  & {\scriptsize YJR113C} &  & {\scriptsize Pxa1} &  & {\scriptsize YPL147W} & {\scriptsize X} &  &  &  & \tabularnewline
{\scriptsize Vps25} &  & {\scriptsize YJR102C} & {\scriptsize X} & {\scriptsize Mrp1} &  & {\scriptsize YDR347W} &  & {\scriptsize Pxa2} &  & {\scriptsize YKL188C} & {\scriptsize X} &  &  &  & \tabularnewline
\cline{1-4} \cline{9-12} 
\multicolumn{4}{l}{} & {\scriptsize Rsm25} &  & {\scriptsize YIL093C} &  &  &  &  &  &  &  &  & \tabularnewline
\multicolumn{4}{l}{} & {\scriptsize Nam9} &  & {\scriptsize YNL137C} &  &  &  &  &  &  &  &  & \tabularnewline
\cline{5-8} 
\end{tabular}
\par\end{centering}

\textbf{a.} Numbers in parentheses refer to the ID of the complex,
if available, from \url{http://yeast-complexes.embl.de} \cite{GAG+06}.
Compositions and localizations of complexes were also taken from references
listed in square brackets. Symbols: {}``{*}'' the protein was not
localized in the compartment \cite{HFG+03}; {}``X'' or {}``NA''
not tagged or no abundance \cite{GHB+03}; {}``a'', ''b'', etc.
refer to outliers in Fig. \ref{fig:cmplx_loga}.
\end{table}

The model complexes used in this study are identified in Table \ref{table:complexes}.
Each complex was nominally associated with a subcellular compartment
based on the names and descriptions of the complexes available in
the literature. Some exceptions are the cyclin-dependent protein kinase
complex, the proteins of which are largely cytoplasmic and nuclear
\cite{HFG+03}, but here is placed in the slot for the ambiguous location
because no definitely ambiguously localized complexes could be identified.
For a similar reason, the proteins listed in Table \ref{table:complexes}
under punctate composite are not part of a named complex but were
chosen because they are localized to early Golgi in addition to the
punctate composite characterization \cite{HFG+03}. Other exceptions
are the vacuolar model proteins (proteases and other canonical vacuolar
proteins \cite{SCC+07}), enzymes of the ergosterol biosynthetic pathway,
some of which are associated with the lipid particle \cite{MB05},
and proteins integral to the peroxisomal membrane, which were identified
using the Gene Ontology (GO) annotations in the SGD \cite{SGD}. Where
they could be found, the ID numbers of the complexes in a yeast complex
database \cite{GAG+06} are listed in parentheses in Table \ref{table:complexes},
as are literature references that describe the composition and/or
localization of the complexes. If any of the proteins in the complexes
do not localize \cite{HFG+03} to the compartment shown in Table \ref{table:complexes}
they are marked with an asterisk; those proteins that were not present
in the YeastGFP database or that are lacking an abundance count therein
\cite{GHB+03} are marked with {}``X'' and {}``NA'', respectively.

\begin{figure}
\begin{centering}
\includegraphics[width=0.95\columnwidth]{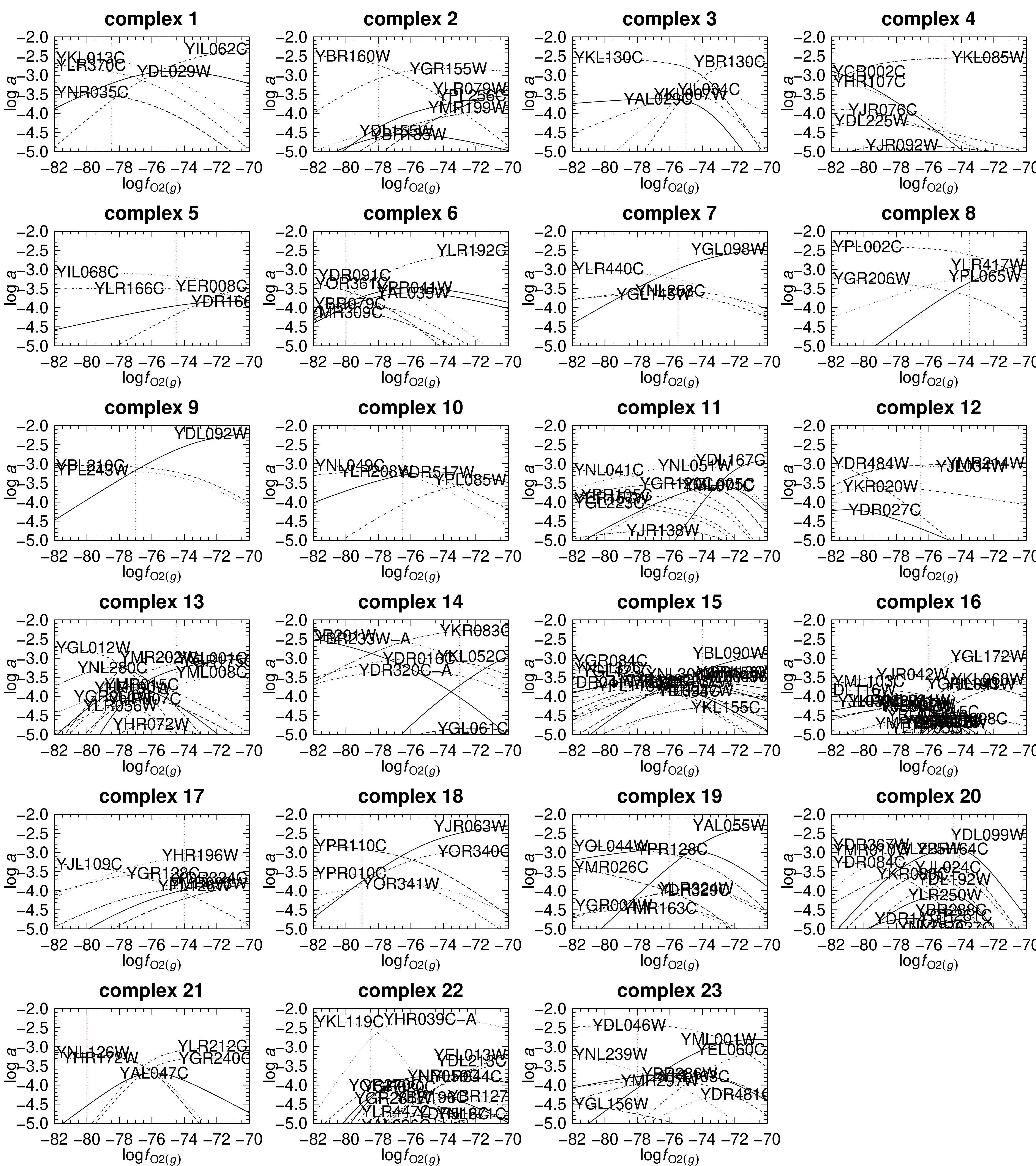}
\par\end{centering}

\caption{\label{fig:cmplx_logfO2}\textbf{Calculated logarithms of activities
of model proteins in complexes}. The numbered complexes are identified
in Table \ref{table:complexes}. Metastable equilibrium activities
of proteins in the complexes were calculated as a function of $\log f_{\mathrm{O_{2}}_{\left(g\right)}}$
for total activity of residues set to unity. Dotted red lines denote
values of $\log f_{\mathrm{O_{2}}_{\left(g\right)}}$ (listed in Table
\ref{table:fO2}) and calculated relative abundances that were used
in making Fig. \ref{fig:cmplx_loga}.}

\end{figure}

\begin{figure}
\begin{centering}
\includegraphics[width=0.95\columnwidth]{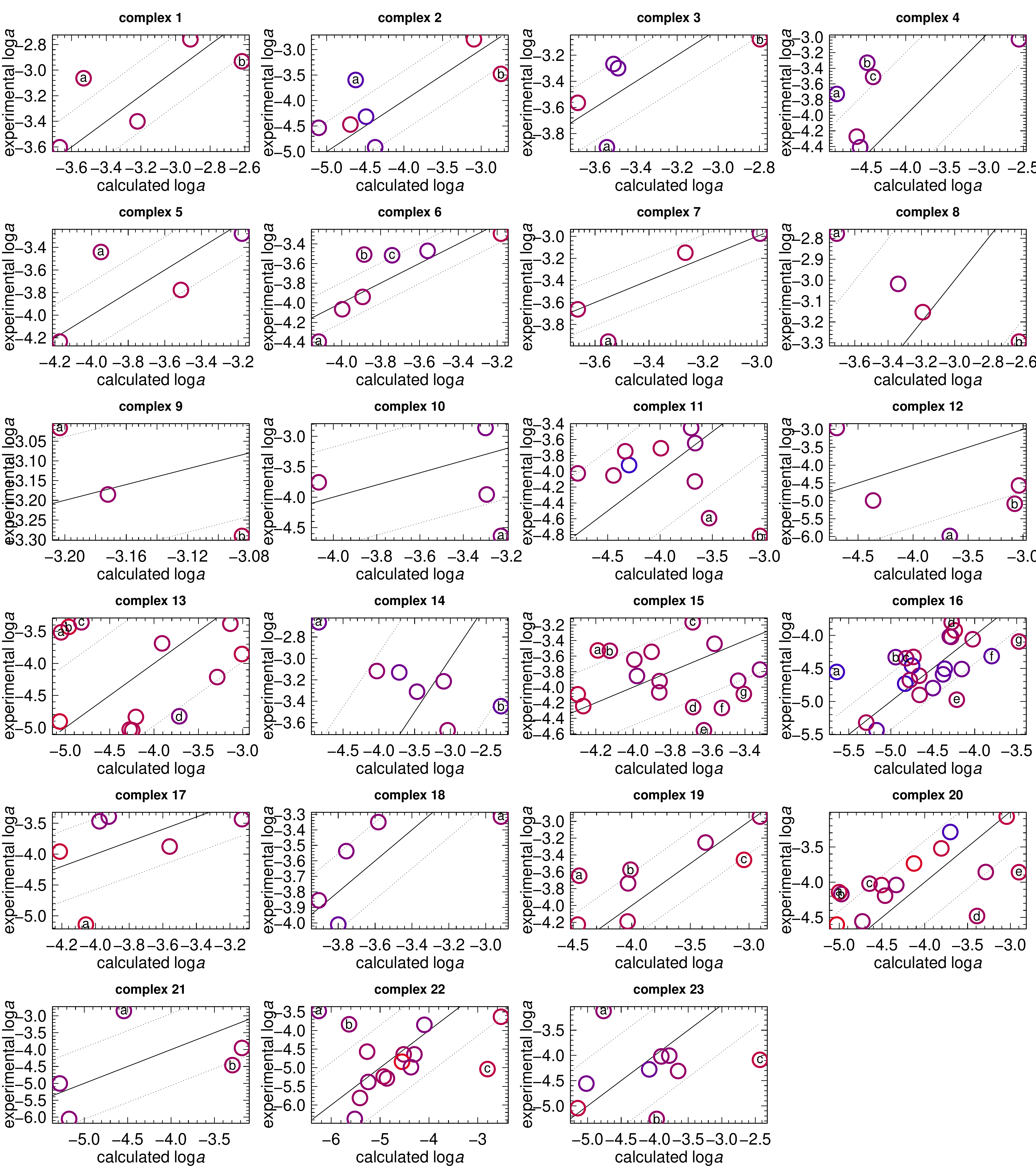}
\par\end{centering}

\caption{\label{fig:cmplx_loga}\textbf{Comparison of experimental and calculated
logarithms of activities of interacting proteins}. Symbols are as
in Fig. \ref{fig:proteome_loga}; the model proteins and the outliers
are listed in Table \ref{table:complexes}.}

\end{figure}

The calculated metastable equilibrium logarithms of activities of
the proteins in each complex are shown as a function of $\log f_{\mathrm{O_{2}}_{\left(g\right)}}$
in Fig. \ref{fig:cmplx_logfO2}. The calculated logarithms of activities
of the proteins were compared with experimental ones by constructing
scatterplots at $0.5\log$ unit intervals from $\log f_{\mathrm{O_{2}}_{\left(g\right)}}=-82$
to $-70.5$, which are shown in Figure S1. As above, visual assessment
of fit was the first resort to obtain values of $\log f_{\mathrm{O_{2}}_{\left(g\right)}}$
that maximize the correspondence with experimental relative abundances,
but the RMSD and Spearman rank correlation coefficient were also considered
in these comparisons. Because of the small sample size in many of
the comparisons, the sign of the correlation coefficient is as useful
as its magnitude in assessing the results. The resulting calculated
relative abundances are listed together with the experimental ones
in Table S4.

The number of model proteins in each of the complexes is less than
the number of most abundant proteins in each compartment considered
in the preceding section, so the visible decrease in scatter is expected.
Some of the model complexes represented in Fig. \ref{fig:cmplx_loga}
exhibit an apparent positive correlation between calculated and experimental
logarithms of activities; these include translation initiation factor
eIF3, nuclear pore complex and proteins integral to peroxisomal membrane.
An inverse correlation between calculated and experimental logarithms
of activities is apparent for proteins in the ESCRT I \& II complexes,
signal recognition complex, and DASH complex. A few of the other complexes
(Golgi transport complex, sterol biosynthesis enzymes) exhibit very
little overall correspondence between calculated and experimental
logarithms of activities.

The results in Fig. \ref{fig:cmplx_loga} permit an interpretation
of the relative energetic requirements for formation of different
groups of interacting proteins. Take for example complex 14, which
is the DASH complex that associates with the microtubule. An inverse
correlation between the experimental and calculated relative abundances
is apparent for this complex in Fig. \ref{fig:cmplx_loga}. The RMSD
between calculated and experimental logarithms of activities of proteins
is $1.05$, which is among the highest listed in Table \ref{table:fO2}.
Note from Eqn. (\ref{eq:affinity}) that a $\sim1$ log unit change
in the chemical activity of a chemical species corresponds to a Gibbs
energy difference equal to $2.303RT$. An average difference of $\sim1$
between calculated and experimental logarithms of activity indicates
that the formation of the proteins requires $2.303RT=1364$ cal mol$^{-1}$
per protein beyond what would be needed if the proteins formed in
metastable equilibrium relative abundances. On the other hand, the
formation in specific oxidation-reduction conditions of proteins making
up translation initiation factor eIF3 and other assemblages where
cellular abundances positively correlate with and span the same range
as the metastable equilibrium distribution can proceed close to a
local minimum energy required for protein formation.

Because of their relatively high energy demands, proteins in complexes
such as the DASH complex and the spindle pole body are likely to be
more dynamic in the cell. Although a positive rank correlation coefficient
for the latter complex is reported in Table \ref{table:fO2}, at a
higher oxygen fugacity ($\log f_{\mathrm{O_{2}}_{\left(g\right)}}=-76$)
a strong inverse correlation obtains between experimental abundances
and calculated metastable equilibrium relative abundances of the proteins
in this complex (Figure S1). The finding made elsewhere of some inverse
relationships between relative abundance of proteins and corresponding
mRNA levels was also interpreted as evidence for additional effort
on the part of the cell \cite{TKR07}. An inverse relationship that
opposes equilibrium may be favored in evolution because of the strategic
advantage of incorporating otherwise costly (rare) amino acids that
increase enzymatic diversity \cite{Wic87}. The present results show
that specific examples of inverse relationships in the relative abundances
of proteins can be identified using a metastable equilibrium reference
state that is conditioned by oxidation-reduction conditions. Chemical
selectivity in the dynamic formation in the cell of high-energy proteins
could lead to transient formation of complexes that function only
under certain conditions. In contrast, complexing proteins that interact
close to metastable equilibrium are more likely to be constitutively
formed.

\section*{Concluding Remarks}

This study was concerned with thermodynamic selectivity of protein
formation primarily as a function of one variable: oxidation-reduction
potential represented by the logarithm of the fugacity of oxygen ($\log f_{\mathrm{O_{2}}_{\left(g\right)}}$).
In reality, many variables are changing in cells, including the hydration
state, $\mathrm{pH}$, activity of $\mathrm{CO_{2}}$ and $\mathrm{H_{2}S}$,
temperature and pressure. These all factor into the Gibbs energy changes
accompanying the overall chemical transformation between proteins.
Except for oxygen fugacity, the other variables were held constant
in most of the calculations reported here. It is tempting to explore
the effects of these variables on the compositions of metastable equilibrium
assemblages. Incorporation into the framework of protein folding reactions
and a non-ideality contribution, or excess Gibbs energy, that would
encompass the effects of electrostatic interactions and macromolecular
crowding is another target for expanding the scope of the thermodynamic
characterizations.

The model results reported above were chosen in order to test specific
predictions made using the hypothesis that the selection for or against
metastable equilibrium has measurable consequences in organisms. The
findings can be summarized as:
\begin{enumerate}
\item The oxidation-reduction potential ($\log f_{\mathrm{O_{2}}_{\left(g\right)}}$)
limits of relative metastabilities of redoxin isoforms overlap with
measured $\mathrm{Eh}$ (redox potential) in the cytoplasm and mitochondrion
but not the nucleus.
\item The model proteologs represent the overall amino acid compositions
of proteins in different compartments. At relatively low oxidation-reduction
potential, proteologs in order of decreasing relative metastability
are those of ER, Golgi, cell periphery, mitochondrion, nuclear periphery
and spindle pole. At higher oxidation-reduction potential, proteologs
in order of decreasing relative metastability are those of actin,
nucleolus, nucleus, vacuole, bud neck and microtubule. At intermediate
oxygen fugacities, proteologs of lipid particle, peroxisome and early
Golgi are relatively metastable compared to those of cytoplasm, vacuolar
membrane and late Golgi.
\item In a chemically reacting system starting with the cytoplasmic proteolog
where all interactions include the proteologs of microtubule or spindle
pole, environmental shifts in $\log f_{\mathrm{O_{2}}_{\left(g\right)}}$
going from $-75$ to $-83.5$ to $-69$ to $-75$ can drive the sequential
formation of proteologs of spindle pole, microtubule, cytoplasm, actin,
nucleus, cell periphery, bud neck and bud.
\item Oxidation-reduction potentials within $-78<\log f_{\mathrm{O_{2}}_{\left(g\right)}}<-74$
give rise to metastable equilibrium populations of most abundant model
proteins within compartments in which the range of protein abundance
becomes closest to that seen in reported measurements. Substantial
scatter is evident in the comparisons, but a moderate overall positive
rank correlation was observed.
\item Closer fits between calculated and experimental relative abundances
were obtained within $-80<\log f_{\mathrm{O_{2}}_{\left(g\right)}}<-73$
by considering fewer numbers of model proteins that interact in complex
formation. Strong positive correlations were found for, among others,
cytoplasmic translation initiation factor eIF3 and nuclear pore complex;
negative correlations were found for the microtubule-associated DASH
complex and the endosomal ESCRT I \& II complexes.
\end{enumerate}
This study contributes to understanding the products of evolution
by quantifying the extent of departure from metastable equilibrium
in populations of interacting proteins. The observed positive correlations
are consistent with a trend of some populations of interacting proteins
to be imprinted with the consequences of local energy minimization
in chemical reactions. These results and observations also support
the notion that changing oxidation-reduction potential can selectively
promote or hold back the reactions leading to formation of complexing
proteins in relative abundances seen in the cell. Combining proteomic
data with metastable equilibrium calculations is therefore a promising
avenue for predicting complexes that form in specific oxidation-reduction
conditions that vary temporally and spatially in biochemical systems.

\section*{Methods}

The essential steps in the calculations reported here are 1) defining
standard states, 2) identifying model proteins for systems of interest,
3) assessing the relative abundances of model proteins in metastable
equilibrium, 4) visualizing the results of the calculations on speciation
or predominance diagrams and 5) comparing the computational results
with experimental biochemical and proteomic data.

\subsection*{Standard states and chemical activities}

The activity of a species is fundamentally related to the chemical
potential of the species by\begin{equation}
\mu=\mu^{\circ}+RT\ln a\,,\label{eq:mu}\end{equation}
where $R$ and $T$ represent, respectively, the gas constant and
the temperature, $\mu$ and $\mu^{\circ}$ stand for the chemical
potential and standard chemical potential, respectively, and $a$
denotes activity. No provision for activity coefficients of proteins
or other species was used in this study; under this approximation,
the activity of an aqueous species is equal to its concentration (molality).

The standard state for aqueous species including proteins specifies
unit activity of the aqueous species in hypothetical one molal solution
referenced to infinite dilution. The standard molal Gibbs energies
of the proteins were calculated with the CHNOSZ software package \cite{Dic08a}
using group additivity properties and parameters taken from \cite{DLH06}.

\subsection*{Proteologs: overall compositions of proteins in compartments}

The overall amino acid compositions of proteins in 23 subcellular
locations in \emph{S. cerevisiae} were calculated by combining localization
\cite{HFG+03} and abundance \cite{GHB+03} data for proteins measured
in the YeastGFP project with amino acid compositions of proteins downloaded
from the \emph{Saccharomyces} Genome Database (SGD) \cite{SGD}. Of
4155 ORF names listed in the YeastGFP dataset, all but 12 are present
in SGD (the missing ones are YAR044W, YBR100W, YDR474C, YFL006W, YFR024C,
YGL046W, YGR272C, YJL012C-A, YJL017W, YJL018W, YJL021C and YPR090W).

To generate proteologs that are most representative of each compartment,
proteins that were annotated in the YeastGFP study as being localized
to more than one compartment were excluded from this analysis (except
for bud; see below), as were those for which no abundance was reported.
The names of the open reading frames (ORFs) corresponding to the proteins
in the YeastGFP data set were matched against the SGD's \texttt{protein\_properties.tab}
file downloaded on 2008-08-04. This search yielded a number of model
proteins for each compartment, ranging from 5 (ER to Golgi) to 746
(cytoplasm); see Table \ref{table:proteologs}. The names of the compartments
used throughout the tables and figures in this paper correspond to
the notation used in the YeastGFP data files (where spaces are replaced
with a period).

It was found that no proteins with reported abundances and localized
to the bud were exclusive to that compartment, hence all of the proteins
localized there (which also have localizations in other compartments)
were taken as models for the bud proteolog. The amino acid composition
of the proteolog for each compartment was calculated by taking the
sum of the compositions of each model protein for a compartment in
proportion to its fractional abundance in the total model protein
population of the compartment. The resulting amino acid compositions
are listed in Table S1. The corresponding chemical formulas of the
nonionized proteologs and the calculated standard molal Gibbs energies
of formation from the elements at 25 $^{\circ}$C and 1 bar of the
ionized proteologs are shown in Table \ref{table:proteologs}.

\subsection*{Metastability calculations}

Diagrams showing the predominant proteins and the relative abundances
of proteins in metastable equilibrium were generated using the CHNOSZ
software package \cite{Dic08a}. These calculations take account of
formation reactions of the proteins written for their residue equivalents
\cite{Dic08a}. This approach is demonstrated in the Results for a
specific model system. 

The basis species, or perfectly mobile components of an open system
\cite{Kor66}, appearing in the formation reactions studied here are
$\mathrm{CO_{2}}_{\left(aq\right)}$, $\mathrm{H_{2}O}$, $\mathrm{NH_{3}}_{\left(aq\right)}$,
$\mathrm{O_{2}}_{\left(g\right)}$, $\mathrm{H_{2}S}_{\left(aq\right)}$
and $\mathrm{H^{+}}$. The reference activities used for the basis
species were $10^{-3}$, $10^{0}$, $10^{-4}$, $10^{-7}$ and $10^{-7}$,
respectively, for $\mathrm{CO_{2}}_{\left(aq\right)}$, $\mathrm{H_{2}O}$,
$\mathrm{NH_{3}}_{\left(aq\right)}$, $\mathrm{H_{2}S}_{\left(aq\right)}$
and $\mathrm{H^{+}}$. In the case of diagrams showing $\mathrm{Eh}$
as a variable, the aqueous electron ($e^{-}$) was substituted for
$\mathrm{O_{2}}_{\left(g\right)}$ in the basis species. Reference
values for $a_{\mathrm{e^{-}}}$ or $f_{\mathrm{O_{2}}_{\left(g\right)}}$
are not listed here because one or the other is used as an independent
variable in each of the calculations described above.

\subsection*{Conversion between scales of oxidation-reduction potential }

Conversion between the $\log f_{\mathrm{O_{2}}_{\left(g\right)}}$
and $\mathrm{Eh}$ scales of oxidation-reduction potential can be
made by first writing the half-cell reaction for the dissociation
of $\mathrm{H_{2}O}$ as\begin{equation}
\mathrm{H_{2}O}\rightleftharpoons\frac{1}{2}\mathrm{O_{2}}_{\left(g\right)}+2\mathrm{H^{+}}+2e^{-}\,.\label{react:water}\end{equation}
Taking $\mathrm{pH}=-\log a_{\mathrm{H^{+}}}$ and $\mathrm{pe}=-\log a_{e^{-}}$,
the logarithmic analog of the law of mass action for Reaction \ref{react:water}
can be written as\begin{equation}
\log K_{\ref{react:water}}=\frac{1}{2}\log f_{\mathrm{O_{2}}_{\left(g\right)}}-2\mathrm{pH}-2\mathrm{pe}-\log a_{\mathrm{H_{2}O}}\,,\label{eq:water.logK}\end{equation}
where $\log K_{\ref{react:water}}$ stands for the logarithm of the
equilibrium constant of Reaction \ref{react:water} as a function
of temperature and pressure. $\mathrm{Eh}$ is related to $\mathrm{pe}$
by \cite{Dre97} \begin{equation}
\mathrm{pe}=\frac{F}{2.303RT}\mathrm{Eh}\,,\label{eq:pe-Eh}\end{equation}
 where $F$ and $R$ denote the Faraday constant and the gas constant,
respectively. Combining Eqns. (\ref{eq:water.logK}) and (\ref{eq:pe-Eh})
yields the following expression for $\mathrm{Eh}$ as a function of
$\log f_{\mathrm{O_{2}}_{\left(g\right)}}$ and other variables:\begin{equation}
\mathrm{Eh}=\frac{2.303RT}{F}\left(\frac{1}{2}\log f_{\mathrm{O_{2}}_{\left(g\right)}}-2\mathrm{pH}-\log a_{\mathrm{H_{2}O}}-\log K_{\ref{react:water}}\right)\,.\label{eq:Eh-logfO2}\end{equation}
At 25$^{\circ}$C and 1 bar, $F/2.303RT=16.903$ volt$^{-1}$ and
$\log K_{\ref{react:water}}=-41.55$; for $\mathrm{pH}=7$ and $\log a_{\mathrm{H_{2}O}}=0$,
a value of $\mathrm{Eh}=0$ V corresponds to $\log f_{\mathrm{O_{2}}_{\left(g\right)}}=-55$.
Eqn. (\ref{eq:Eh-logfO2}) permits the conversion between $\mathrm{Eh}$
and $\log f_{\mathrm{O_{2}}_{\left(g\right)}}$ as well at other temperatures,
$\mathrm{pH}$s, and activities of $\mathrm{H_{2}O}$.

\subsection*{Average nominal oxidation state of carbon}

Let us write the chemical formula of a species of interest as $\mathrm{C}_{n_{\mathrm{C}}}\mathrm{H}_{n_{\mathrm{H}}}\mathrm{N}_{n_{\mathrm{N}}}\mathrm{O}_{n_{\mathrm{O}}}\mathrm{S}_{n_{\mathrm{S}}}^{Z}$,
where $Z$ denotes the net charge. The average nominal oxidation state
of carbon ($\overline{Z}_{\mathrm{C}}$) of this species is given
by\begin{equation}
\overline{Z}_{\mathrm{C}}=\frac{Z-n_{\mathrm{H}}+2\left(n_{\mathrm{O}}+n_{\mathrm{S}}\right)+3n_{\mathrm{N}}}{n_{\mathrm{C}}}\,.\label{eq:ZC}\end{equation}
Eqn. (\ref{eq:ZC}) is consistent with the electronegativity rules
described in \cite{HCH70} and is compatible with the equation for
average oxidation number of carbon used in \cite{Buv83}. For example,
Eqn. (\ref{eq:ZC}) can be used to calculate the average nominal oxidation
states of carbon in $\mathrm{CO_{2}}$ and $\mathrm{CH_{4}}$, which
are $+4$ and $-4$, respectively. Note that the proportions of oxygen
and other covalently-bonded heteroatoms contribute to the value of
$\overline{Z}_{\mathrm{C}}$ of a protein or other molecule, but that
proton ionization does not alter the nominal carbon oxidation state,
because of the opposite contributions from $Z$ and $n_{\mathrm{H}}$
in Eqn. (\ref{eq:ZC}). In the 4143 proteins identified in the YeastGFP
subcellular localization study and found in the \emph{Saccharomyces}
Genome Database, the minimum and maximum of $\overline{Z}_{\mathrm{C}}$
are $-0.414$ and $0.390$, respectively. Of the proteins in this
dataset, six have $\overline{Z}_{\mathrm{C}}<-0.35$ (YDR193W, YDR276C,
YEL017C-A, YJL097W, YML007C-A, YMR292W) and six have $\overline{Z}_{\mathrm{C}}>0.15$
(YCL028W, YHR053C, YHR055C, YKR092C, YMR173W, YPL223C). The points
in the scatterplots in this paper (Figs. \ref{fig:proteome_loga}
and \ref{fig:cmplx_loga} and Figure S1) are colored on a continuous
red-blue scale according to the value of $\overline{Z}_{\mathrm{C}}$
of the proteins, where maximum red occurs at $\overline{Z}_{\mathrm{C}}=-0.35$
and maximum blue occurs at $\overline{Z}_{\mathrm{C}}=0.15$.

\subsection*{}

\subsection*{Comparison with experimental relative abundances}

In comparison, experimental abundances of proteins in each model system
were scaled so that the total chemical activity of residues was equal
to unity.

The root mean square deviation between calculated and experimental
logarithms of activities was calculated using

\begin{equation}
\mathrm{RMSD}=\sqrt{\frac{\sum_{i=1}^{n}\left(X_{\mathrm{calc},i}-X_{\mathrm{expt},i}\right)^{2}}{n}}\,,\label{eq:RMSD}\end{equation}
where $X_{\mathrm{calc},i}$ and $X_{\mathrm{expt},i}$ denote the
calculated and experimental logarithms of activities and $n$ stands
for the number of proteins.x

The Spearman rank correlation coefficient ($\rho$) was calculated
using\begin{equation}
\rho=1-\frac{6d}{n\left(n^{2}-1\right)}\,,\label{eq:spearman}\end{equation}
where $d=\sum_{i=1}^{n}\left(x_{\mathrm{calc},i}-x_{\mathrm{expt},i}\right)^{2}$
and $x_{\mathrm{calc},i}$ and $x_{\mathrm{expt},i}$ stand for the
ranks of the corresponding logarithms of activities.

\section*{Supporting Information}

\subsection*{Figure S1: Comparisons of relative abundances of proteins (PDF)}

Scatterplots of experimental vs. calculated abundance ranking and
logarithm of activity of most abundant proteins and selected complexes
in subcellular compartments are shown as a function of oxygen fugacity.

\subsection*{Table S1: Amino acid compositions of model proteologs (CSV)}

Overall amino acid compositions of proteins in subcellular locations
of \emph{S. cerevisiae} were calculated from YeastGFP localization
\cite{HFG+03} and abundance \cite{GHB+03} data downloaded from \url{http://yeastgfp.ucsd.edu/}
combined with protein compositions downloaded from the \emph{Saccharomyces}
Genome Database (\url{http://www.yeastgenome.org/}). The amino acid
compositions of the proteologs were used to calculate the properties
listed in Table \ref{table:proteologs}.

\subsection*{Table S2: Intercompartmental protein reactions (TXT)}

This table lists chemical reactions between residue equivalents of
proteologs for interacting compartments. The charges of the proteologs
were calculated at 25 $^{\circ}$C, 1 bar and $\mathrm{pH}=7$.

\subsection*{Table S3: Abundance data for model proteins in compartments (CSV)}

For the up to 50 most abundant model proteins in each compartment
are listed the ORF name, sequence length, average nominal oxidation
state of carbon (Eqn. \ref{eq:ZC}), computed standard molal Gibbs
energy at 25 $^{\circ}$C and 1 bar of the ionized protein and charge
at $\mathrm{pH=7}$ and calculated and experimental logarithm of activity.

\subsection*{Table S4: Abundance data for protein complexes (CSV)}

For the model complexes in each compartment (see Table \ref{table:complexes})
are listed the same properties as in Table S3.

\subsection*{Text S1: CHNOSZ software package (GZ)}

This is the complete package (source code, documentation and data
files) for the CHNOSZ program, which was used together with the program
script (below) to perform the calculations in this study. The package
is designed to be used with the R software environment \url{http://www.R-project.org}.
Additional information about CHNOSZ is available in \cite{Dic08a}
and at \url{http://www.chnosz.net}.

\subsection*{Text S2: Program script and data files for generating figures (GZ)}

This program script and supporting files were used to generate the
figures shown above. It includes the script itself (plot.R), protein
compositions (generated from the protein\_properties.tab file downloaded
from the \emph{Saccharomyces} Genome Database), calculated standard
molal thermodynamic properties of the proteins (to speed up calculations),
YeastGFP protein localization and abundance data \cite{HFG+03,GHB+03},
and a .csv version of Table \ref{table:fO2}. To generate the figures,
the contents of the zip file should all be placed into the R working
directory before loading CHNOSZ. Then read in the script with \texttt{source('plot.R')}.
More details on the operation are provided at the top of the script
file.

\subsection*{Text S3: Interactions between subcellular compartments in yeast (PDF)}

This file lists statements from \cite{BAM+97,KGS97,JWH97,WGC97,LK97,PS91}
used to identify the interactions between proteins in different compartments
of \emph{Saccharomyces cerevisiae} that are listed in Table \ref{table:interactions}.

\bibliographystyle{plos}
\bibliography{aa}

\end{document}